%% file: main.tex
\documentclass[lettersize,journal]{IEEEtran}
\usepackage{amsmath,amsfonts,amssymb}
\usepackage{algorithmic}
\usepackage{array}
\usepackage[caption=false,font=normalsize,
   labelfont=sf,textfont=sf]{subfig}
\usepackage{textcomp}
\usepackage{stfloats}
\usepackage{url}
\usepackage{verbatim}
\usepackage{graphicx}
\usepackage{balance}
\usepackage{algorithm}
\usepackage{siunitx}
\usepackage{xurl}
\usepackage{subcaption}
\usepackage{hyperref}
\usepackage{cleveref}
\usepackage{xcolor}
\usepackage{tabularx}
\usepackage{booktabs}
\usepackage{csquotes}
\usepackage{comment}
\usepackage{microtype}
\usepackage[style=ieee,doi=true,isbn=false,url=false, backend=biber]{biblatex}

\AtEveryBibitem{\clearlist{language}}
\newcolumntype{C}[1]{>{\centering\arraybackslash}m{#1}}

\input{tikz/setup.tex}
\newcommand{\mrm}[1]{\mathrm{#1}}

\begin{document}

\title{Complex-Phase, Data-Driven Identification of Grid-Forming Inverter Dynamics}
\author{Anna~Büttner,~\IEEEmembership{Non~Member,~IEEE,}
        Hans~Würfel,~\IEEEmembership{Non~Member,~IEEE,}
        Sebastian~Liemann,~\IEEEmembership{Member,~IEEE,}
        Johannes~Schiffer,~\IEEEmembership{Member,~IEEE,}
        and ~Frank~Hellmann,~\IEEEmembership{Non~Member,~IEEE,} 
\thanks{A. Büttner, H. Würfel and F. Hellman are with the Potsdam Institute for Climate Impact Research, Potsdam, Germany, e-mail: \{buettner,wuerfel,hellmann\}@pik-potsdam.de}
\thanks{S. Liemann was with the Technical University Dortmund, Dortmund, Germany, e-mail: sebastian.liemann@tu-dortmund.de}
\thanks{J. Schiffer is with the Brandenburg University of Technology Cottbus-Senftenberg and the Fraunhofer IEG, Fraunhofer Institution for Energy Infrastructures and Geothermal Energy Systems, Cottbus, Germany, e-mail: schiffer@b-tu.de}
}
\maketitle

\newcommand{\dq}{\mrm{dq}}
\newcommand{\ifilt}{i_\mrm{f}}
\newcommand{\iout}{i_\mrm{g}}
\newcommand{\uout}{v_\mrm{C}}

\newcommand{\Pmeas}{P^\mrm{m}}
\newcommand{\Qmeas}{Q^\mrm{m}}
\newcommand{\Pref}{P^\mrm{s}}
\newcommand{\Qref}{Q^\mrm{s}}
\newcommand{\Pfilt}{\bar P}
\newcommand{\Qfilt}{\bar Q}
\newcommand{\Vref}{v^\mrm{s}}
\newcommand{\imeas}{i^\mrm{m}}
\newcommand{\Vmeas}{v^\mrm{m}}
\newcommand{\wref}{\omega^\mrm{s}}
\newcommand{\cf}{\eta}
\newcommand{\cfpred}{\dot{\Theta}^\mrm{pred}}
\newcommand{\cfmeas}{\dot{\Theta}^\mrm{m}}
\newcommand{\cp}{\Theta}
\newcommand{\cppred}{\Theta^\mrm{pred}}
\newcommand{\cpmeas}{\Theta^\mrm{m}}
\newcommand{\xc}{x_\mrm{c}}
\newcommand{\numeas}{\nu^\mrm{m}}
\newcommand*\diff{\mathop{}\!\mathrm{d}}

\begin{abstract}
    The increasing integration of renewable energy sources (RESs) into power systems requires the deployment of grid-forming inverters to ensure a stable operation. Accurate modeling of these devices is necessary. In this paper, a system identification approach to obtain low-dimensional models of grid-forming inverters is presented. The proposed approach is based on a Hammerstein-Wiener parametrization of the normal-form model. The normal-form is a gray-box model that utilizes complex frequency and phase to capture non-linear inverter dynamics. The model is validated on two well-known control strategies: droop-control and dispatchable virtual oscillators. Simulations and hardware-in-the-loop experiments demonstrate that the normal-form accurately models inverter dynamics across various operating conditions. The approach shows great potential for enhancing the modeling of RES-dominated power systems, especially when component models are unavailable or computationally expensive.
\end{abstract}



\section{Introduction} 
    \IEEEPARstart{R}{enewable} energy sources (RESs) and the power-electronic inverters that connect them to the grid play an increasingly important role for the electric power mix. Today, most inverters utilize grid-following strategies and require an external grid for synchronization. These inverters face challenges in maintaining grid stability under low shares of synchronous generation \cite{kroposki_achieving_2017}. In contrast, grid-forming inverters can offer functionalities that have been provided by synchronous generators, including frequency and voltage control, independently \cite{matevosyan_grid-forming_2019}. 
    
    Grid-forming inverters are vital for the safe operation of RES-dominated systems, which has led scientists and transmission grid operators to prioritize the development and integration of grid-forming inverters~\cite{german_transmission_system_operators_need_2020, matevosyan_grid-forming_2019}. Given their significance in RES-dominated power grids, grid-forming inverters must be modeled appropriately.
    
    Often, commercially available inverters have to be treated as black boxes, as manufacturers do not fully disclose the details of their internal design \cite{chakraborty_review_2022, helman_grey-box_2024}. Without access to models, there are limited insights into how these devices behave in interconnected systems. Operating grid-forming inverters without this knowledge introduces risks and makes it harder to anticipate and mitigate potential failures.
     
    Another challenge is that existing, disclosed models tend to be computationally intensive due to detailed cascaded control structures. The computational effort limits the number of fault scenarios that can be analyzed, rendering these models impractical for control-room applications such as dynamic stability assessment tools~\cite{matevosyan_grid-forming_2019}. Therefore, identifying low-dimensional models from data is crucial. These models may enable stability assessments without comprehensive knowledge of the underlying design and with feasible computational effort.

    The review paper \cite{chakraborty_review_2022} provides an excellent overview of the state-of-the-art in system identification for power systems. While it explicitly highlights the importance of system identification approaches for obtaining mathematical models of inverter-based resources, it also shows that the current research has been focused on grid impedance identification. Although impedance equivalents play a crucial role in the small signal-stability analysis of grid-forming inverters, they inherently assume a linear transfer function. As a result, they cannot capture nonlinear effects or accurately describe system behavior under faults.

    In recent years, there has been increasing focus on data-driven modeling approaches that capture nonlinear dynamics \cite{helman_grey-box_2024, qi_synchronization_2022, guruwacharya_data-driven_2024, abdelsamad_voltage-source_2021}. The authors of \cite{helman_grey-box_2024} propose a grey-box modeling approach; however, it simplifies the system to a purely linear time-invariant (LTI) representation, considering only frequency dynamics while entirely neglecting voltage magnitude. While \cite{qi_synchronization_2022} incorporates both frequency and voltage dynamics, its primary focus is on synchronization stability rather than providing an accurate dynamic model. Similarly, \cite{abdelsamad_voltage-source_2021} proposes a nonlinear identification approach focused on estimating harmonic characteristics, but only for the steady state. In \cite{guruwacharya_data-driven_2024} a nonlinear system identification approach is presented, but it is limited to grid-forming inverters operating under droop control—an assumption also made in \cite{qi_synchronization_2022, helman_grey-box_2024}. 

    Our approach, in contrast, is fully data-driven and nonlinear. It does not rely on a priori droop characteristics and allows for full dynamic simulations. We demonstrate that it is applicable not only to droop-controlled inverters but also to other grid-forming control strategies, such as dispatchable virtual oscillator control.
    
    In this work, we build upon the normal-form modeling framework for grid-forming inverters introduced in \cite{kogler_normal_2022}, which provides a theory-driven gray-box model class capable of capturing key nonlinearities in a unified formulation. While \cite{kogler_normal_2022} introduced the model class and demonstrated its applicability in a proof of concept, it did not develop a systematic identification approach based on time-series data, which will be introduced in this work.

    Inspired by the recent success of the complex frequency concept~\cite{milano_complex_2022}, in \cite{buttner_complex_2024}, the connection between the normal-form model and the complex frequency \cite{milano_complex_2022} has been established.

    Extending this work, we now introduce the complex phase, derived from the complex frequency, which will be presented in detail in Section~\ref{sec:complex_phase}. For the first time, we demonstrate that the normal-form model admits a so-called Hammerstein-Wiener (H-W) parametrization, characterized by nonlinear input and output transformations, with a linear subsystem that determines the system dynamics. This insight, detailed in Section~\ref{sec:nf}, allows us to design a tailored identification pipeline that leverages the H-W structure and uses the complex phase as a target variable for the identification.

    The focus of this paper is twofold. First, we introduce the novel system identification pipeline, which targets the complex phase, detailed in Section~\ref{sec:parameter_identification}. Second, we utilize this pipeline to identify the H-W normal-form for two types of grid-forming inverters: those based on droop-control~\cite{schiffer_conditions_2014} and those based on the dispatchable virtual oscillator~(dVOC)~\cite{colombino_global_2017, seo_dispatchable_2019}. These inverters are described in detail in Section~\ref{sec:devices_under_test}. The identification of the dVOC-inverter is based on a Simulink EMT-model introduced in~\cite{tayyebi_frequency_2020}. The droop-controlled inverter is identified based on experimental data measured in a power hardware-in-the-loop laboratory~\cite{krishna_power-hardware---loop_2022}. 
        
    The data-collection process and the data requirements are outlined in section~\ref{sec:data_collection}. The major advantage of the approach introduced here is that it is purely data-driven, enabling model identification for inverters without available models, such as the laboratory inverter.

    While this paper focuses on identifying models for grid-forming inverters, the approach can also be applied to other grid-forming components, such as static synchronous compensators with appropriate control strategies.

    The results are presented in Section~\ref{sec:results}. It has been demonstrated that the normal-form succeeds at describing the slow dynamics of both inverters. The ability of the normal-form to capture the inverter dynamics across various operating conditions is showcased. The normal-form has also been implemented in Simulink to study its performance in a closed-loop set-up. Finally, Section~\ref{sec:limitations} discusses the limitations and potential solutions of the approach.

     All relevant software and data needed to reproduce the results presented here, including the identified system matrices, test case parameters, and measured time series, are available on Zenodo \cite{buttner_software_2024}.
    
    \section{Theoretical Background}
    \subsection{Notation}
    A three-phase voltage signal $V_a$, $V_b$, $V_c$ is considered balanced if it satisfies $V_a + V_b + V_c = 0$ at all times. Using the Clarke and Park transformations \cite{machowski_power_2008}, such balanced signals can be represented in a co-rotating 2D reference frame, known as $\mathrm{dq}$-coordinates. In this paper, we consider all voltages and currents balanced, thus transform all $\mrm{abc}-$signals into a common \emph{global} $\dq$-frame of constant synchronous frequency.
    The $\dq$-voltage is represented as a complex number with phase $\varphi$ and magnitude $V \geq 0$, i.e.,
    \begin{equation}
        v(t) = v_d(t) + j\,v_q(t) = V(t)\, e^{j \varphi(t)}. \label{eq:dq_voltage}
    \end{equation}
    Currents are transformed analogously to voltages. The complex power $S$ is defined as:
    \begin{align}
        S = P + j Q = v\,i^* \label{eq:power}
    \end{align}
    where $P$ and $Q$ are the active and reactive power respectively, and $i^*$ is the complex conjugate of the current $i$.
    
    \subsection{System Identification}
    System identification refers to the process of developing mathematical models of dynamical target systems based on measured data. In this study, the target systems are grid-forming inverters. System identification always deals with input-output relationships. Inputs are the external signals applied to the target system, while outputs are the measured responses resulting from these inputs. In this work, we utilize gray-box identification. Gray-box modeling leverages partial knowledge of the system behavior and completes the model using empirical data, e.g. extracting model parameters from measurements.
    
    Selecting an appropriate gray-box model necessitates an understanding of the underlying target system. Grid-forming inverters control the voltage and frequency at their output. Hence, fundamentally, we assume that grid-forming inverters can be modeled as controllable voltage sources. The following generic model can emulate the behavior of such a controllable voltage source:
    \begin{equation}
    \begin{aligned}
        \dot{x}(t) &= h^x(x, i), \\
        v(t) &= h^v(x), \label{eq:controlled_voltage_source}
    \end{aligned}
    \end{equation}
    where $v$ and $i$ are the inverter $\mathrm{dq}$-voltages and currents, respectively, $x$ is a set of internal model states and $h^x$ and $h^v$ are the, possibly, non-linear mappings to be learned. 
    
    From an engineering perspective, a natural consideration for inputs and outputs for the system identification of \eqref{eq:controlled_voltage_source} would be the current $i$ and voltage $v$. However, the relationship between voltages and currents in grid-forming inverters is non-linear and non-stationary, which complicates the identification process. Motivated by the modeling approach for grid-forming actors given in \cite{kogler_normal_2022}, we address this issue by using a different set of inputs and outputs that, based on the results reported in this paper, seems to be more suitable for system identification. 
    
    \subsection{Complex Phase and Frequency}
    \label{sec:complex_phase}
    The currently accepted definition of frequency in power grids is meaningful only when the voltage magnitude is constant. Otherwise, this definition fails to separate the effects of variations in phase angle and voltage magnitude. The concept of complex frequency, introduced by F.~Milano in~\cite{milano_complex_2022}, addresses this limitation by incorporating the dynamics of both the phase angle $\varphi$ and the voltage magnitude $V$. 
    The complex frequency $\cf$ can be derived by expressing the complex $\mathrm{dq}$-voltage $v(t)$, as defined in equation \eqref{eq:dq_voltage}, as
    \begin{align}
        v(t) &= e^{\operatorname{ln}(V) + j \varphi} = e^{\Theta}\qquad\text{with}\label{eq:voltage_complex_phase} \\
        \cf &= \rho + j \omega \label{eq:complex_frequency}\,,
    \end{align}
    where the imaginary part of the complex frequency, $\omega$, represents the angular frequency, and the real part, $\rho$, denotes the relative rate of change of the voltage magnitude. Here, $\Theta$ is referred to as the \emph{complex phase}. The complex frequency framework provides a definition of frequency that applies to all balanced conditions, including transients. The complex frequency concept has already been applied in various contexts, including inertia estimation of virtual power plants~\cite{zhong_-line_2022}.

    In this paper, we employ a universal modeling approach for grid-forming devices based on complex frequency and phase to model inverter dynamics.
    
\begin{figure}
        \centering
        \includegraphics[width=\linewidth]{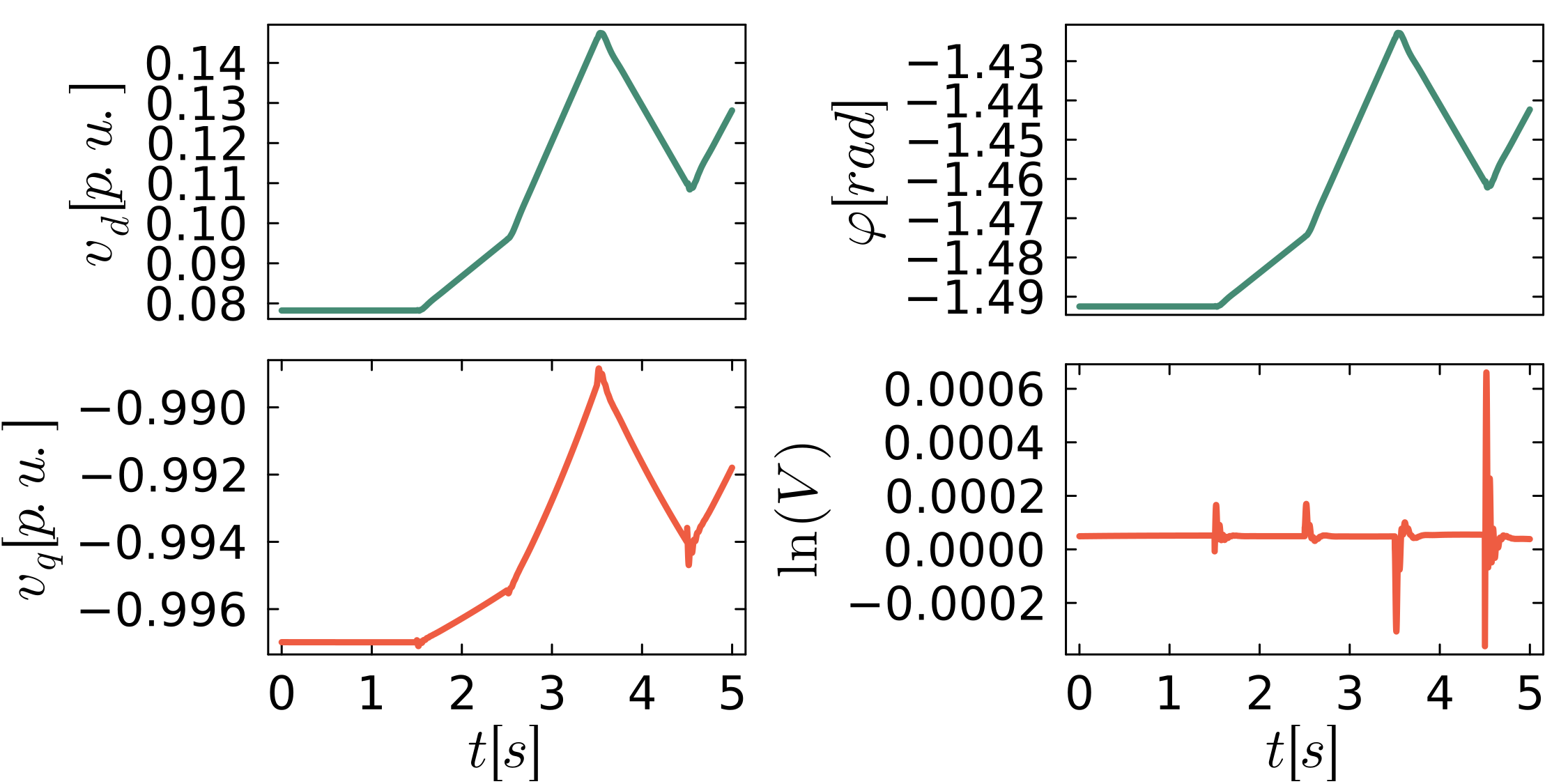}
        \caption{Comparison between $\mathrm{dq}$-voltage and the complex phase transients.}
        \label{fig:dq_voltage_complex_phase}
    \end{figure}

    While both the complex phase and the $\mathrm{dq}$-voltages contain identical information, the complex phase is more suitable for identification purposes as it separates phase and magnitude variations. In Fig.~\ref{fig:dq_voltage_complex_phase} we compare exemplarily $\mathrm{dq}$-voltage and complex phase transients. The $\mathrm{d}$-component and phase angle $\varphi$ exhibit similar behavior. The $\mathrm{q}$-component and voltage logarithm $\operatorname{ln}(V)$ both display small, rapid oscillations following a fault. However, the $\mathrm{q}$-component is additionally overlaid with a slower signal, which complicates the identification.
    While the separation of magnitude and phase dynamics is also achieved by writing the voltage in polar coordinates it is not well suited for the modeling approach, which will be introduced in the following section.

    \subsection{Normal-Form of Grid-Forming Devices} 
    \label{sec:nf}
    Kogler et al. \cite{kogler_normal_2022} introduced the \textit{normal-form}, a technology-neutral, gray-box model of grid-forming devices. The normal-form is derived by employing a set of natural assumptions. For an in-depth introduction to these assumptions and the derivation of the model, readers are referred to the original publication~\cite{kogler_normal_2022}.
    
\begin{figure}[H]
        \centering
        \input{tikz/normalform_lti.tex}
        \caption{A single grid-forming inverter, modeled by the normal-form, coupled to the power grid.}
        \label{fig:normal-form}
    \end{figure}
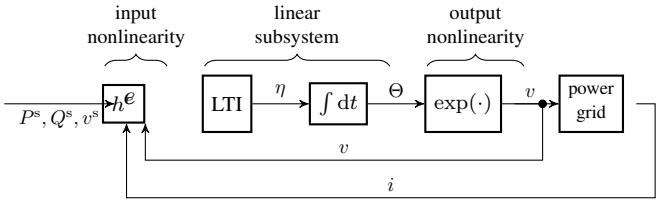

    The main result of the normal-form approach is that the dynamics of the complex phase $\cp$ only depend on the active and reactive power $P$ and $Q$, calculated via equation \eqref{eq:power}, the voltage magnitude square $\nu$ and a set of internal model states $\xc$. The quantities $P$, $Q$, and $\nu$ are considered with their respective error coordinates $e$ relative to the set-points $\Pref$, $\Qref$, and $\Vref$:
    \begin{align}
        e &= h^e\left(v, i, P^s, Q^s, \nu^s\right)\, , \label{eq:err} \\
        &= \left(\Re( v\,i^*), \Im(v\,i^*), |v|^2\right)^T - \left(\Pref, \Qref, |\Vref|^2\right)^T. 
    \end{align}
    This yields the following most universal form of the normal-form model:
    \begin{equation}
      \begin{aligned}
        \dot{x}_\mrm{c}(t) &= g(e, \xc)\,,\\
       \cf(t) &= f(e, \xc)\,, \\
       \dot{\Theta}(t) &= \cf \,, \\
       v(t) &= e^{\Theta} \,, \\
      \end{aligned} \label{eq:nf}
    \end{equation}
    where $f$ and $g$ are continuous, non-linear functions. If $f$ and $g$ are linear, the normal-form exhibits a Hammerstein-Wiener (H-W) structure. H-W models are characterized by a static non-linearity at the input, followed by a linear subsystem that defines the system dynamics, and ending with a non-linearity that computes the output \cite{ljung_system_1998}. In the normal-form, the input non-linearity is given by equation \eqref{eq:err}. The output non-linearity is defined by the computation of the $\mathrm{dq}$-voltage via the complex phase \eqref{eq:voltage_complex_phase}. The H-W normal form is defined as:
    \begin{equation}
        \begin{aligned}
        e &= h^e(v, i, P^s, Q^s, \nu^s) \,, \\
        \dot{x}_\mrm{c} &= A \xc + B e\,, \\
        \cf &= C \xc + D e\,, \\
        \dot{\Theta} &= \cf \,, \\
       v &= e^{\Theta} \,. \label{eq:nf_lin}
        \end{aligned}
    \end{equation}
    Note that the matrices $A$ and $B$ are real in this formulation, while $C$ and $D$ are complex. The structure of the H-W normal-form is summarized in Fig.~\ref{fig:normal-form}. In \cite{kogler_normal_2022}, it was analytically demonstrated that the H-W normal form can approximate the control laws of both dVOC- and droop-controlled inverters up to higher order terms in the voltage deviation $\Delta \nu$. This model strictly captures only the main control dynamics, neglecting further control such as those providing the voltage source behavior, which are always present in real inverters. In contrast, we will account for these further control loops in this work.
    
    The H-W version is used in the subsequent analysis, and whenever the normal-form is mentioned, it refers to system \eqref{eq:nf_lin}. To account for the relationship between the inputs $e$ and the voltage $v$, the identification will be based on the complex phase $\Theta$, as detailed in section \ref{sec:opt}. 
    In the normal-form, any structural differences between grid-forming inverters are encapsulated in the parameter matrices $A$, $B$, $C$, and $D$. These parameters can be identified from simulated or measured data. A data-driven normal-form for a simple example has been derived in \cite{kogler_normal_2022}, demonstrating the approach's potential. The number of internal states $n_{ivars}$ of the normal-form is variable and can be adjusted according to the complexity of the inverter dynamics.
    
    \section{Devices under Test}
    \label{sec:devices_under_test}
    To validate the introduced modeling scheme, we find the parameter matrices for two, prominent, independent, grid-forming inverters based on recorded input-output data and assess the prediction performance of the normal-form models.

    \subsection{Dispatchable Virtual Oscillator}
    \label{sec:sim}
\begin{figure}
      \centering
      \input{tikz/emtsim.tex}
      \caption{Cascaded control structure of grid-forming inverter in the EMT simulations. The LCL output filter is a per-phase representation of the actual three-phase filter.}\label{fig:simcontrol}
    \end{figure}
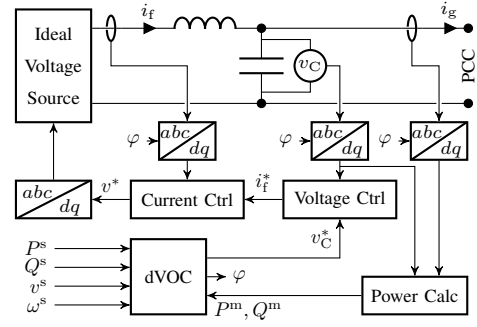

    We identify the normal-form parameters from time series obtained from an EMT simulation to validate the approach in a simulation setting.
    The simulation model consists of a single grid-forming converter using the dVOC control scheme \cite{seo_dispatchable_2019}.
    The Simulink implementation is based on the simulation files from \cite{tayyebi_frequency_2020}.
    In this model, the power electronics are modeled as an ideal voltage source, i.e. the voltage is smooth, and discrete switching events are not considered.
    The power electronics are connected to the grid via an LC-output filter.
    The dVOC algorithm acts as the \emph{outer loop} providing a voltage reference to the \emph{inner loop} controllers.
    The inner loop consists of two cascaded PI controllers in a local $\mathrm{dq}$-frame, which control the inner filter current and voltage.
    The structure of the overall system is shown in Fig.~\ref{fig:simcontrol}.
    For data collection, the inverter is connected to an infinite bus or a resistive load, depending on the scenario.
    The models are available online \cite{buttner_software_2024}, and details on the control design and model parameters can be found in the publications \cite{seo_dispatchable_2019,tayyebi_frequency_2020}.
    
    \subsection{Droop-Controlled Inverter} 
    \label{sec:lab}
\begin{figure}
      \centering
      \includegraphics[trim={0 5 0 35}, clip, width=3.2in]{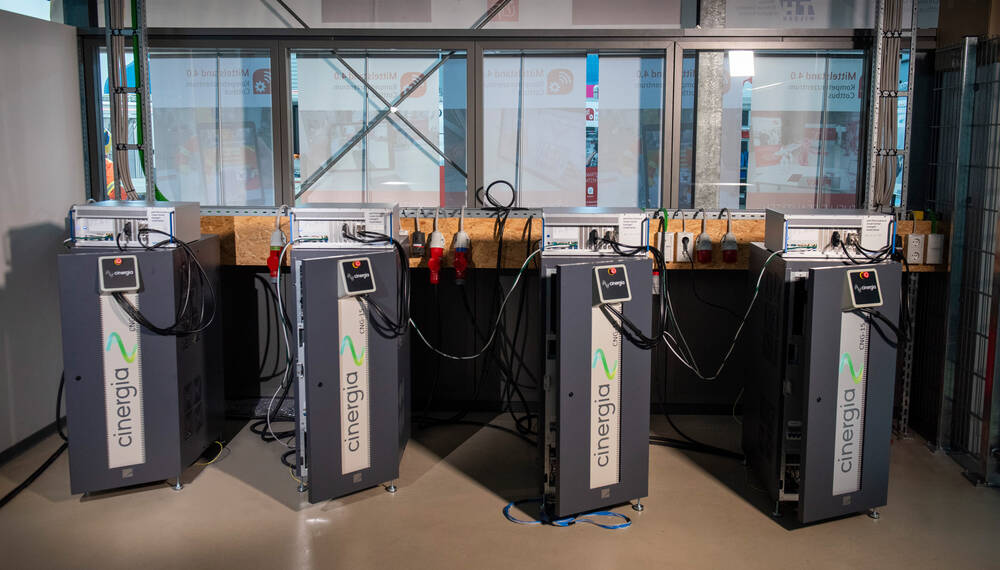}
      \caption{Four inverters in the power hardware in the loop lab. While structurally similar, the implemented control differs significantly from the EMT model.}\label{fig:labphoto}
    \end{figure}

    To validate the system identification pipeline on real data, we gathered experimental data from a grid-forming inverter in the power hardware in the loop laboratory at BTU Cottbus-Senftenberg (see Fig.~\ref{fig:labphoto}).

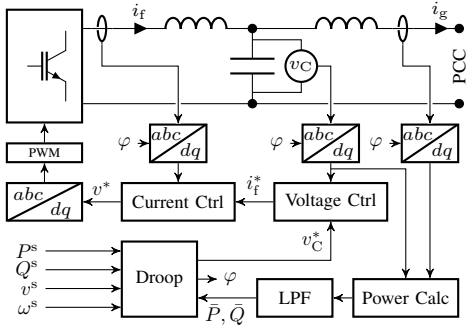
\begin{figure}
      \centering
      \input{tikz/control.tex}
      \caption{Cascaded control structure of grid-forming inverter in the laboratory. The LCL output filter is a per-phase representation of the actual three-phase filter.}\label{fig:labcontrol}
    \end{figure}

    The device under test is a three-phase power converter with a rating of \SI{15}{kW} at a voltage level of \SI{400}{V} phase-to-phase. The implemented grid-forming control strategy is a standard droop control scheme~\cite{schiffer_conditions_2014}, where the desired voltage magnitude $V$ and angle $\delta$ are given by
    \begin{equation*}
      \dot \delta = \wref + K_{\mathrm{P}}\left(\Pref - \Pfilt\right)\quad\text{and}\quad
      V = \Vref + K_{\mathrm{Q}}\left(\Qref - \Qfilt\right),
    \end{equation*}
    where $\Pfilt$ and $\Qfilt$ are the low-pass-filtered active and reactive power  measurements $\Pmeas$ and $\Qmeas$:
    \begin{equation*}
      \tau_{\mathrm{P}} \dot{\Pfilt} = \Pmeas-\Pfilt
      \quad\text{and}\quad
      \tau_{\mathrm{Q}} \dot{\Qfilt} = \Qmeas-\Qfilt\,.
    \end{equation*}
    The inner-loop voltage controller is a two-level cascaded PI-controller in $\mathrm{dq}$-frame with additional resonant terms in the innermost control loop to dampen harmonics in the output.
    The structure of the control loops is presented in Fig.~\ref{fig:labcontrol}. More details on the laboratory and the inner control loops can be found in~\cite{krishna_power-hardware---loop_2022, 
    wurfel_experimentally_2023}.

    \section{Parameter Identification}
    \label{sec:parameter_identification}
    In the following, the system identification pipeline will be introduced. Firstly, the data has to be collected and will then be preprocessed. The identification will be initialized as detailed in sections~\ref{sec:init}. The actual system identification, that is based on an optimization, is outlined in section~\ref{sec:opt}.

    \subsection{Data Collection}
    \label{sec:data_collection}
    To perform the identification, data of the \emph{excited system} needs to be collected. The data collection is specifically tailored to the output, the complex phase, which will facilitate the identification process. The complex phase separates phase and magnitude responses. Consequently, all scenarios are designed to excite both phase and magnitude dynamics, which generates an output with a transient response that is suitable for identification.
    
    For that, a grid-forming inverter is connected to a stiff voltage source. In the EMT simulations, a slack bus has been employed, while in the laboratory a more powerful inverter, that emulates a stiff external grid, has been used. To identify the normal-form parameters, three different scenarios have been measured:
   \begin{enumerate}
        \item Step changes of the voltage magnitude of the external grid voltage ($<$ ±2\% of the nominal voltage)
        \item Step changes of the frequency of the external grid voltage ($<$ ±\SI{50}{mHz}) and
        \item Rapid small changes of both the magnitude and frequency of the external grid voltage ($<$ ±2\% of the nominal voltage and $<$ ± \SI{50}{mHz}).
    \end{enumerate}
    For each of the scenarios, the output voltages $\uout$ and currents $\iout$ are recorded. The two-step change scenarios 2) and 3) have been recorded to capture the inverter's ability to reach one stationary state when starting from another. The small changes have been recorded to study the inverter's response to rapid changes in the external grid without reaching a steady-state between the changes. We found that these three scenarios are sufficient to capture the inverter dynamics fully. 

    The scenarios were designed to be non-destructive and did not trigger any laboratory protection measures, making them suitable for an experimental setting. Further information on the scenarios can be found in the accompanying repository \cite{buttner_software_2024}.The three scenarios recorded here are similar to those used to assess whether an inverter is grid-forming, according to the guidelines introduced in \cite{vde_fnn_fnn_2020}. These guidelines are supported by the German transmission system operators \cite{german_transmission_system_operators_need_2020}.
    
    The data-sets have been partitioned into three categories using a so-called \enquote{train-validation-test} split, a strategy that is commonly employed in the machine learning community. The training set, constituting \SI{70}{\percent} of the total data, is utilized to identify the model parameters. The validation set is used to choose the best parameter configuration. The parameter configuration is selected based on the $R^2$ score, which we will discuss in section~\ref{sec:preformance_measures}, on the validation set. This mitigates over-fitting on the training data. The validation data-set makes up \SI{20}{\percent} of the total data. Subsequently, the test-set evaluates the final performance of the best model, chosen via the validation set, and ensures the model's ability to generalize to unseen data. This data-set comprises \SI{10}{\percent} of the total data. While all three subsets contain different time series, they all include the scenarios introduced above.

    To further challenge the model a so-called \enquote{out-of-distribution} set has been recorded that is significantly different from the training- validation- or test-data. This task is important because the response in real-world power grids can vary widely. It will never be feasible to include all scenarios that may occur in a power grid in the training's data. Hence, a reduced model may only become a viable alternative if it can be shown that it performs well on unseen scenarios.

    In the simulation, the converter is connected to a resistive load instead of the slack, thus acting as a truly grid-forming component. To disturb the system, the resistance of the load decreases multiple times. Therefore, the system is not in power balance anymore, and the frequency drops.
   
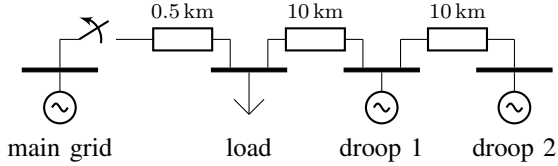
\begin{figure}
         \centering
         \input{tikz/lab_ood.tex}
         \caption{Out of distribution scenario: a micro-grid, containing a load and two grid-forming inverters, disconnects from the main power grid.}
         \label{fig:micro-grid}
     \end{figure}

    In the laboratory, two droop-controlled grid-forming inverters and a constant power load are connected in a micro-grid, which is connected to an auxiliary grid, as shown in figure \ref{fig:micro-grid}. During normal operation, the micro-grid imports active power from the auxiliary grid. In a discrete event, the connection to the auxiliary grid is cut and the grid-forming actors have to compensate for the power loss, which results in a slight frequency drop of the micro-grid.
    
    \subsection{Data Pre-Processing}
    \label{sec:preprosses}
    Before the identification can be performed, the collected data must be preprocessed to fit the normal-form approach. 
    Firstly, the obtained $\mathrm{abc}$ time series of the output current $\iout$ and voltage $\uout$ are transformed into a \emph{global} $\mrm{dq}$-frame rotating with the same synchronous frequency.
    Then, the measured active power $\Pmeas$, reactive power $\Qmeas$, and squared voltage magnitude $\numeas$ are computed from the $\mathrm{dq}$ signals, via the non-linear input transformation \eqref{eq:power}. The difference between the setpoints and the computed values serves as the input for the LTI system \eqref{eq:nf_lin}. The data is down-sampled to a sampling interval of $t_0 = \SI{1}{ms}$ to reduce the computational load as the collected data contains redundant information, especially as we primarily aim to capture slow dynamics that occur on time scales slower than $\sim$\SI{20}{ms}.

    \subsection{Parameter Initialization}
    \label{sec:init}
     A good initial guess for the parameter matrices of the linear subsystems in the H-W models is beneficial to avoid getting trapped in a sub-optimal local minimum during the identification \cite{sjoberg_initializing_2012}. The inputs of the linear subsystem are the error coordinates $e(t)$ and the output is the complex frequency $\cf$. It is possible to apply classical system identification tools to find the system matrices capable of generating the desired complex frequency. We utilized the implementation of the subspace-identification algorithm \cite{ljung_system_1998} available in the \texttt{ControlSystemIdentification.jl} library to calculate the initial parameter matrices $A^0$, $B^0$, $C^0$, and $D^0$. 
    
    The subspace identification algorithm \cite{ljung_system_1998} assumes no feedback from the output back to the input, leading to errors in the predicted complex frequency. Any errors present in the predicted complex frequency accumulate over time during integration, resulting in inaccurate predictions of the $\mathrm{dq}$-voltages. A subspace identification algorithm that accounts for feedback present in our data could be employed, but we have not pursued this, as our goal is to predict the voltages rather than the complex frequency. Measuring the complex frequency $\eta$ directly in the laboratory is impossible and calculating $\eta$ is prone to errors if the measured signals are noisy, as derivatives are inherently sensitive to noise. Therefore, we always focus on predicting the measured voltages. Nevertheless, the initial guess provided by the subspace identification algorithm has proven beneficial, as it allows the optimizer to start from a linearly stable normal-form model.

    \subsection{Parameter Optimization}
    \label{sec:opt}
    The system identification is based on the optimization of the squared $l_2$-norm between the measured complex phase $\cpmeas$ and the predicted complex phase $\cppred$:
    \begin{align}
        l = \sum_k \left(|\cpmeas(t_k) - \cppred(t_k)|\right)^2\,, \label{eq:loss}
    \end{align}
    where $k$ runs over all time points in the time series. For each time step $k$, the LTI-system predicts the complex frequency $\cfpred(t_{k+1})$ based on the error coordinates $e(t_k)$. The complex phase $\cppred (t_{k+1})$ is calculated by integrating the predicted complex frequency.
    
\begin{figure}
        \centering
        \input{tikz/training_flow.tex}
        \caption{Open-loop normal-form identification work-flow.}
        \label{fig:training-flow}
    \end{figure}
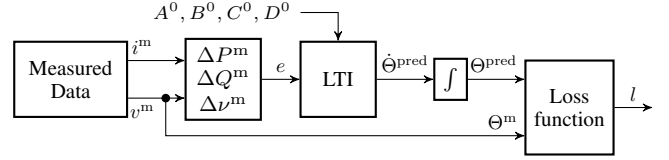

    The parameter matrices $A$, $B$, $C$, and $D$ are adjusted to minimize the loss $l$. For the optimization, the \texttt{Optimization.jl} package \cite{dixit_optimizationjl_2023} and the Julia implementation of the Broyden–Fletcher–Goldfarb–Shanno (BFGS) algorithm \cite{fletcher_practical_2000} have been employed. The full identification work-flow is summarized in Fig.~\ref{fig:training-flow}.
    The inputs $e^m$ for the normal-form are not updated during the identification process, which is the standard procedure for system identification. The inputs depend on the voltages and currents. Consequently, after predicting the voltage $v^{pred}$, updating the currents and the inputs $e$ is necessary for a \enquote{closed-loop} identification. While fixing $e$ may reduce the accuracy, it offers a significant advantage. This approach enables the identification of inverters where no model is available, for example, the laboratory inverter, introduced in section~\ref{sec:lab}. In section~\ref{sec:results_emt}, the normal-form has been implemented in Simulink to test the \enquote{closed-loop} performance, continuously updating the inputs according to the state of the network. This ensures that the identified model is truly accurate.
    
    \subsection{Performance Measures}
    \label{sec:preformance_measures}
    To evaluate the model performance, the $R^2$ score, or coefficient of determination, is used. The $R^2$ score is defined as:
    \begin{align}
        R^2 = 1 - \frac{\sum_i (y_i - f_i)^2}{\sum_i (y_i - \Bar{y})^2},
    \end{align}
    where $y_i$ and $f_i$ represent the observed data points and the corresponding predicted values, respectively. $\Bar{y}$ is the mean of all observed data points. The $R^2$ score ranges from 0 to 1, with 1 indicating a perfect fit and 0 signifying that the model predicts only the mean $\Bar{y}$. In this study, we will consider the $R^2$ scores of the predicted $\mathrm{dq}$ voltages.

    It is important to note that the $R^2$ score can only be used to compare different time series with caution, as the mean $\Bar{y}$ will vary for each scenario, leading to different $R^2$ scores even if the fit quality is similar. In this work, the $R^2$ score will be used primarily to compare the performance of models with different numbers of internal states. A detailed analysis of the measured and predicted voltage transients will be conducted to assess the model performance comprehensively.
    
\section{Results} 
    \label{sec:results}

    \subsection{EMT-Simulations}\label{sec:results_emt}
\begin{figure}
        \centering
        \includegraphics[width=\linewidth]{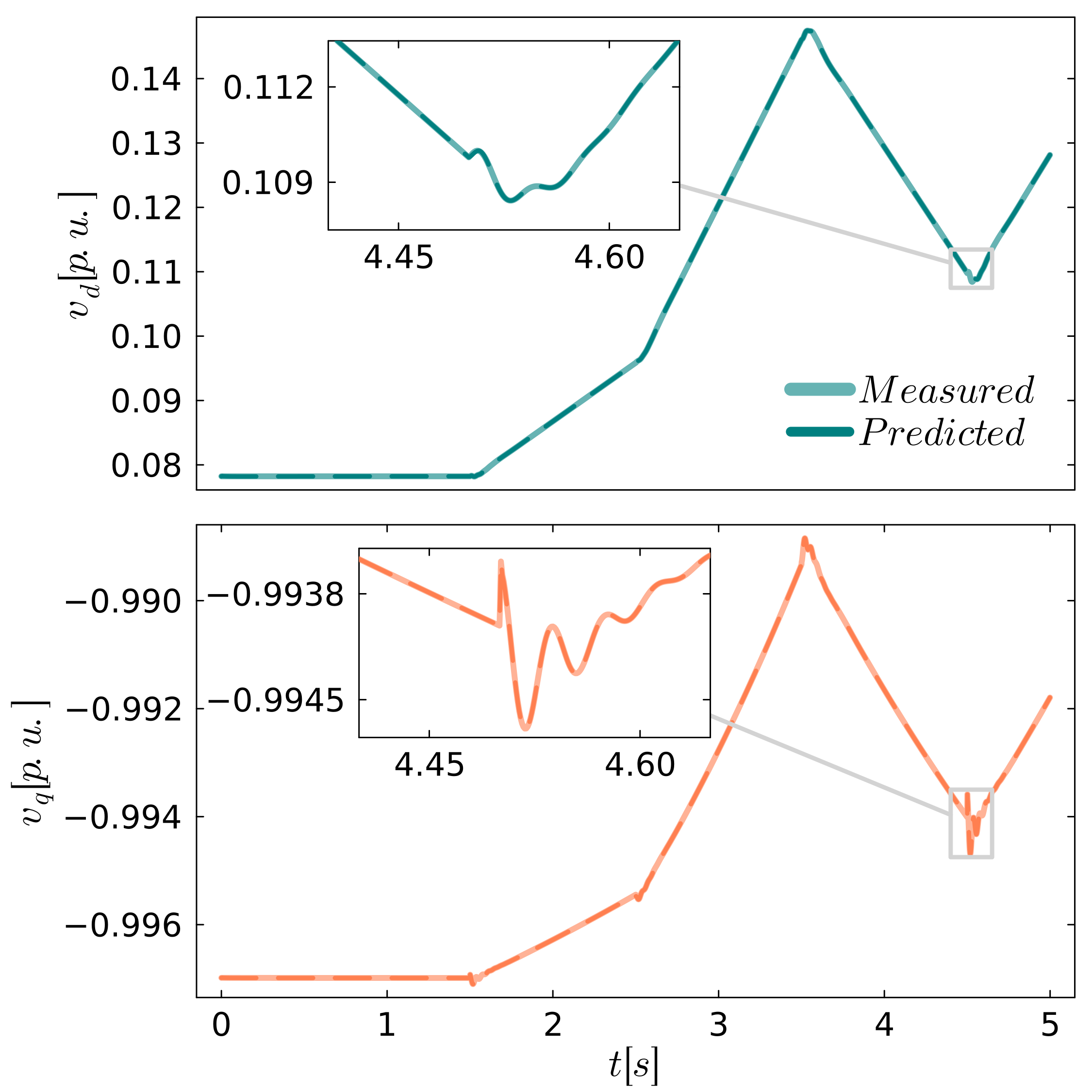}
        \caption{Comparison between the measured and predicted $\mathrm{dq}$-voltages for the third scenario in the test data-set of the dVOC.}
        \label{fig:EMT_test}
    \end{figure}

\begin{figure}
        \centering
        \includegraphics[width=\linewidth]{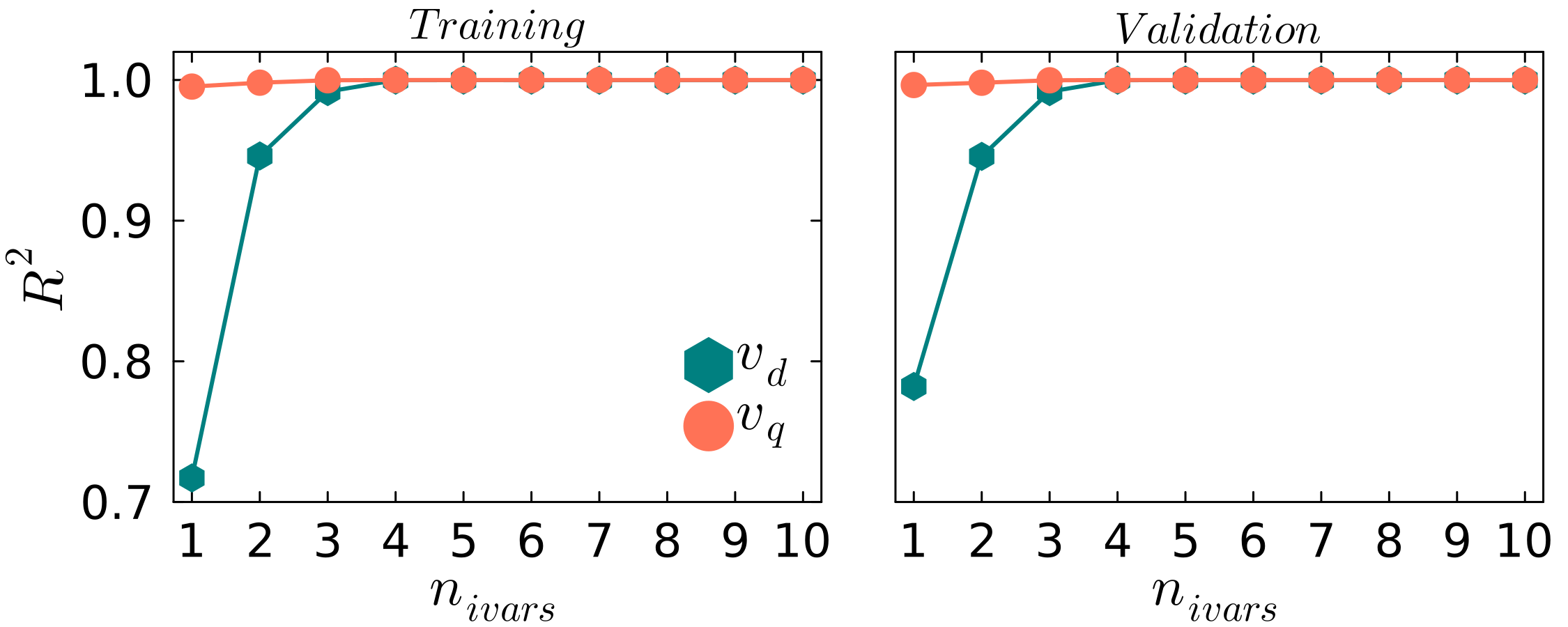}
        \caption{$R^2$ scores with respect to the numbers of internal variables for the dVOC. The plots on the left and right show the results for the training and validation data-set respectively.}
        \label{fig:r2_EMT}
    \end{figure}

    In the subsequent analysis, we present the results for an identified normal-form with four internal variables ($n_{\mathrm{ivars}} = 4$), as Fig.~\ref{fig:r2_EMT} suggests that at least four internal variables are necessary to model the dynamics of the $\mathrm{d}$-component, while the $\mathrm{q}$-component is already captured with a single internal variable. Fig.~\ref{fig:EMT_test} illustrates the results for the third scenario within the test data-set, which involves rapid small changes in both magnitude and frequency. The normal-form successfully captures the voltage dynamics of the inverter in this scenario. The results for the other two scenarios are equally successful, which demonstrates the accuracy of the normal-form model across different operating conditions.

\begin{figure}
        \centering
        \includegraphics[width=\linewidth]{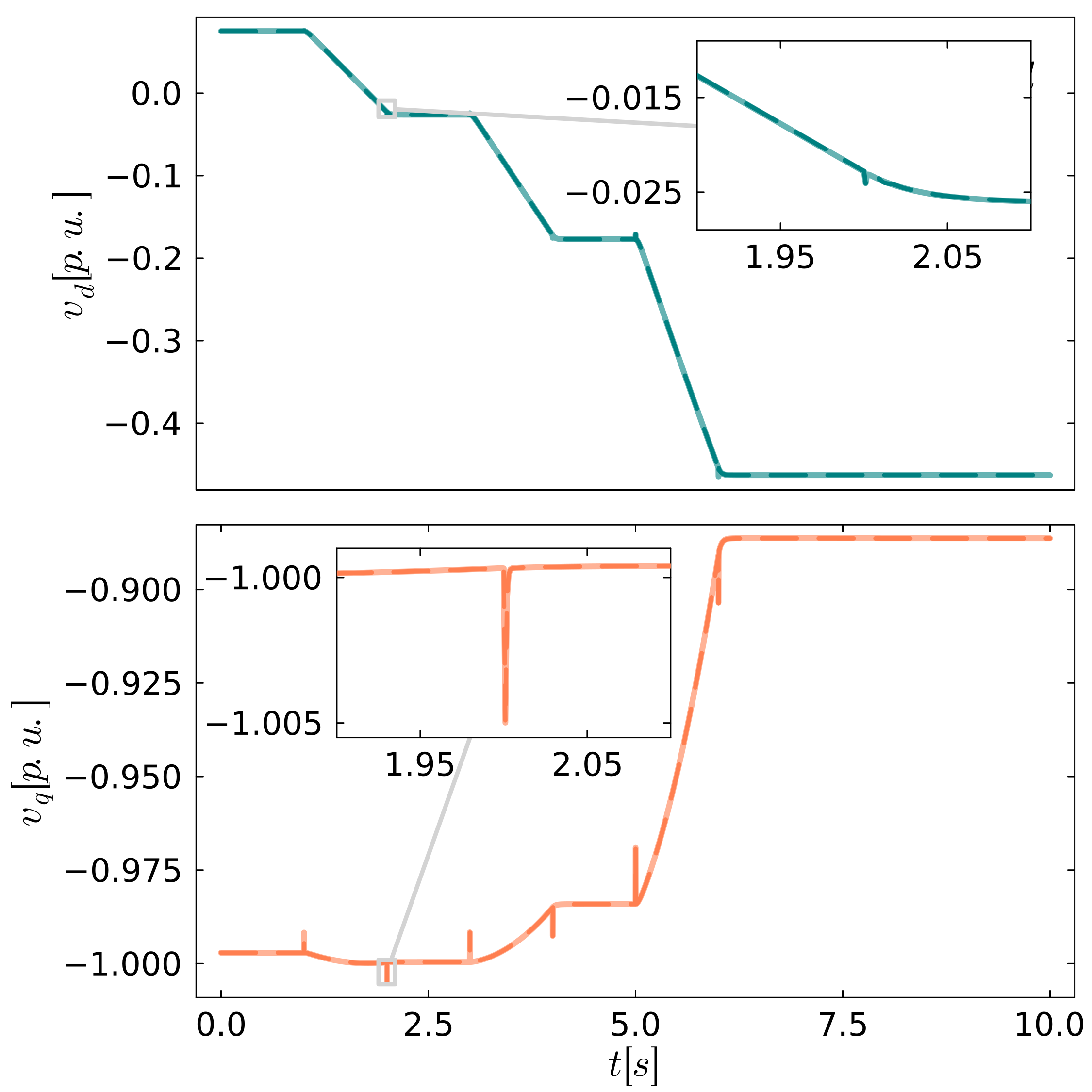}
        \caption{Comparison between the measured and predicted $\mathrm{dq}$-voltages for the load steps scenario of the dVOC.}
        \label{fig:EMT_ood}
    \end{figure}

    As detailed in section~\ref{sec:data_collection}, an out-of-distribution scenario has been collected to assess the ability of the normal-form to predict unforeseen scenarios. Fig.~\ref{fig:EMT_ood} presents a comparison between the predicted and measured $\mathrm{dq}$ -voltage of the out-of-distribution data-set. It reveals a close match between the measured and predicted voltages. It can be seen that the slow transients as well as the fast transients are captured by the normal-form. 
    
    To study the performance with respect to the model complexity, the parameters of the normal-form have been identified for one up to ten internal variables $\xc$. The appropriate number of internal variables is determined by evaluating the $R^2$ score. Specifically, $n_{\mrm{ivars}}$ is increased until adding further variables no longer improves the fit significantly, ensuring that the model captures the system dynamics adequately. The results are depicted in Fig.~\ref{fig:r2_EMT} for both the training- and validation data-set. 
    
\begin{figure}
        \centering
        \includegraphics[width=\linewidth]{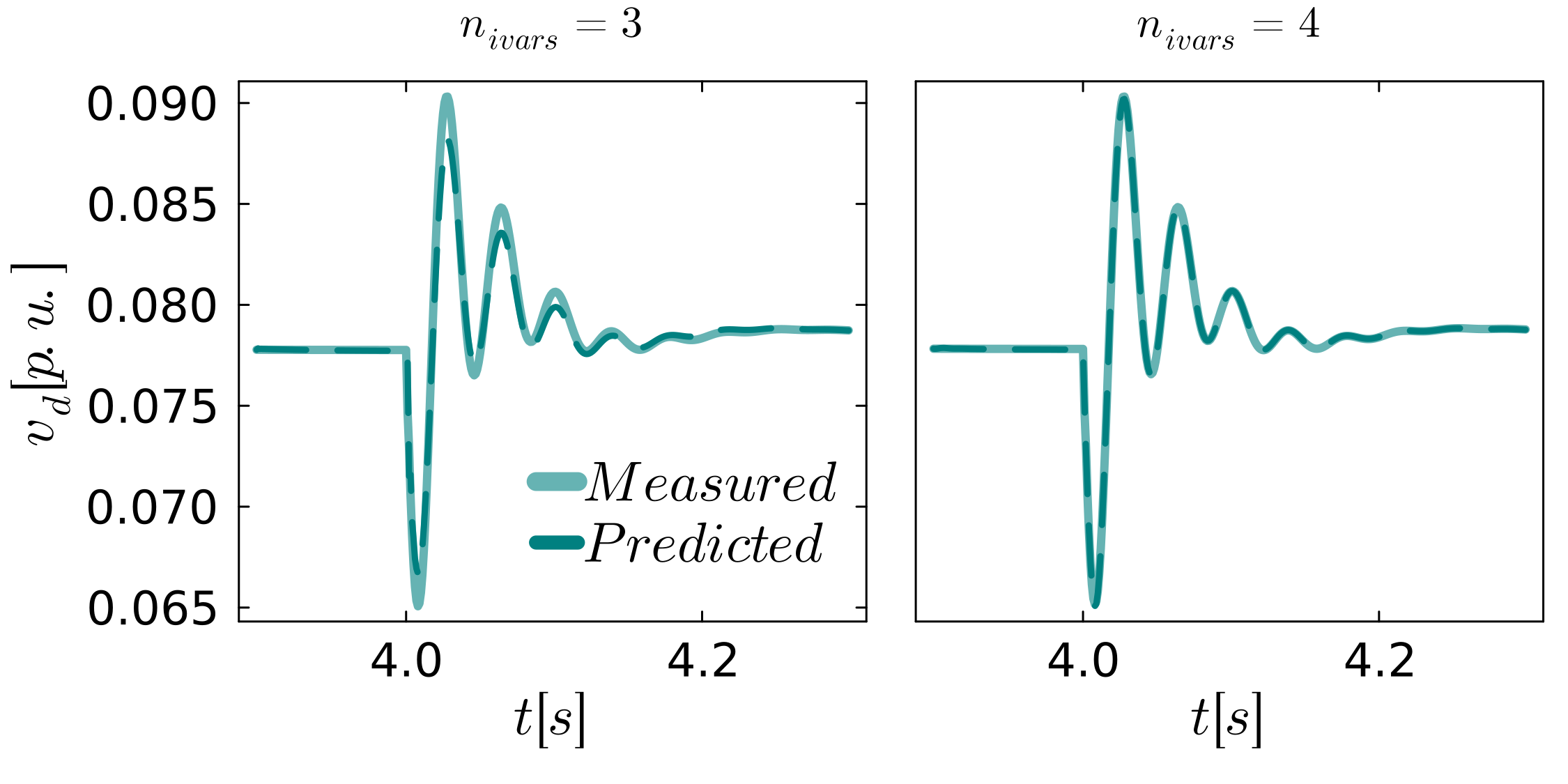}
        \caption{Comparison between normal-form models for the dVOC with three and four internal variables.}
        \label{fig:compare_n_ivars}
    \end{figure}

    Beyond $n_{\mrm{ivars}} = 4$, only marginal improvements of the fit are observed which suggests that the return of increasing the number of internal variables are diminishing. To underline this, Fig.~\ref{fig:compare_n_ivars} shows a comparison between a normal-form with three and four internal variables. It shows voltage transients during a rapid change in slack frequency and magnitude. The normal-form with three internal variables fails to model the $\mathrm{d}$-component, with a high degree of accuracy, while the model with four internal variables successfully captures the dynamics.

    To further validate the approach, the normal-form has been implemented in Simulink to test the model in a closed-loop configuration. As detailed in section~\ref{sec:opt}, the input $y$ remained static during the simulation in previous sections, which is sufficient for training and validation. It is vital to ensure the normal-form performance within a closed-loop context to establish its reliability as an alternative. 
    
\begin{figure}
         \centering
         \includegraphics[width=\linewidth]{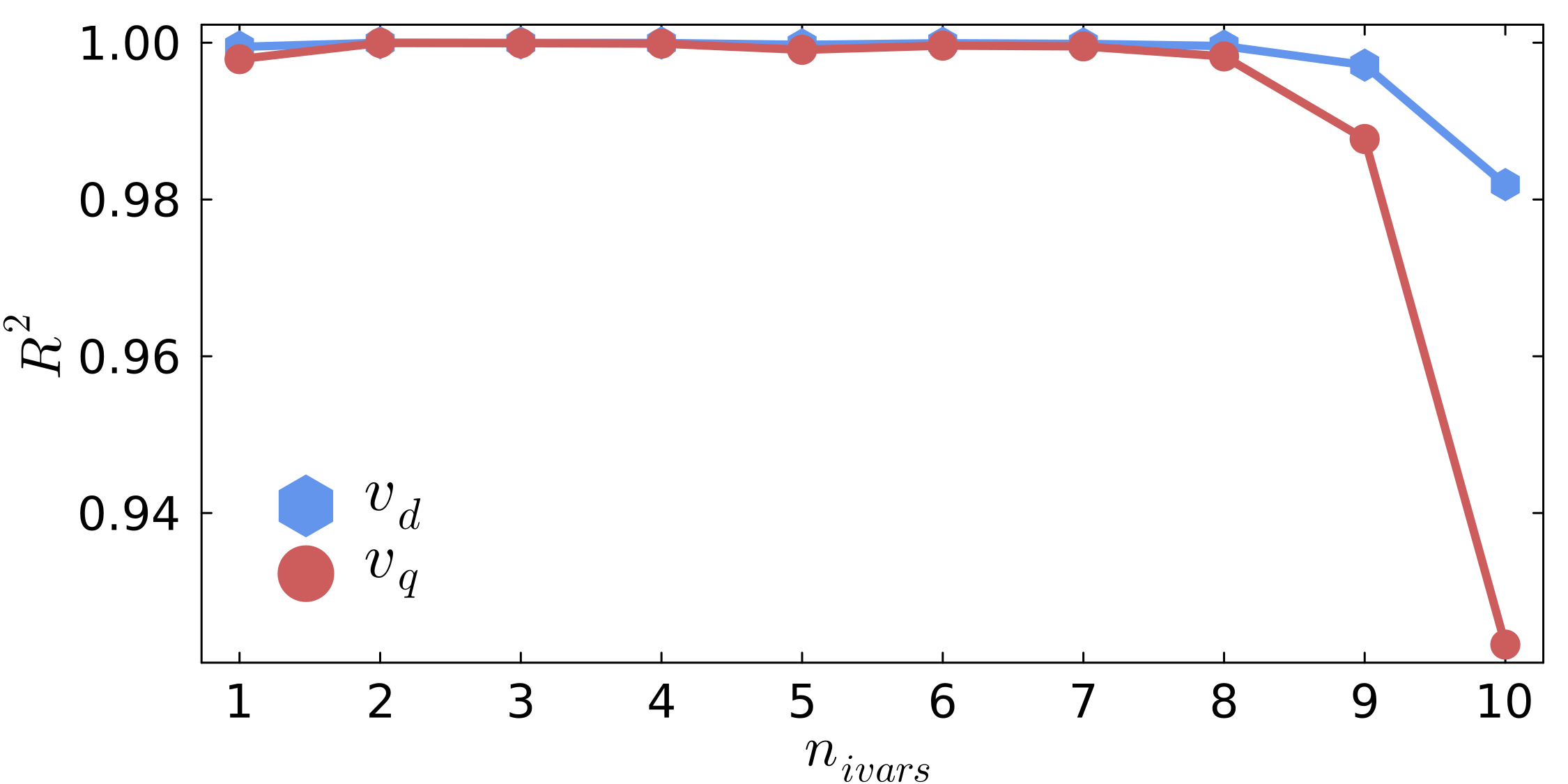}
         \caption{Normal-form closed-loop performance.}
         \label{fig:r2-closed_loop}
    \end{figure}

    Fig.~\ref{fig:r2-closed_loop} illustrates the performance of the closed-loop normal-form in Simulink on the out-of-distribution scenario. In contrast to the open loop performance, depicted in Fig.~\ref{fig:r2_EMT}, it is evident that the models with 1-2 internal variables exhibit a significantly improved performance once the \enquote{back reaction} of the network is considered. The performance remains consistent for models with 3-7 internal variables. However, for 8-10 internal variables, the performance of the models decreases. This loss of performance is likely a result of over-fitting, which may occur for models with a higher number of parameters \cite{lever_model_2016}.

\begin{figure}
         \centering
         \includegraphics[width=\linewidth]{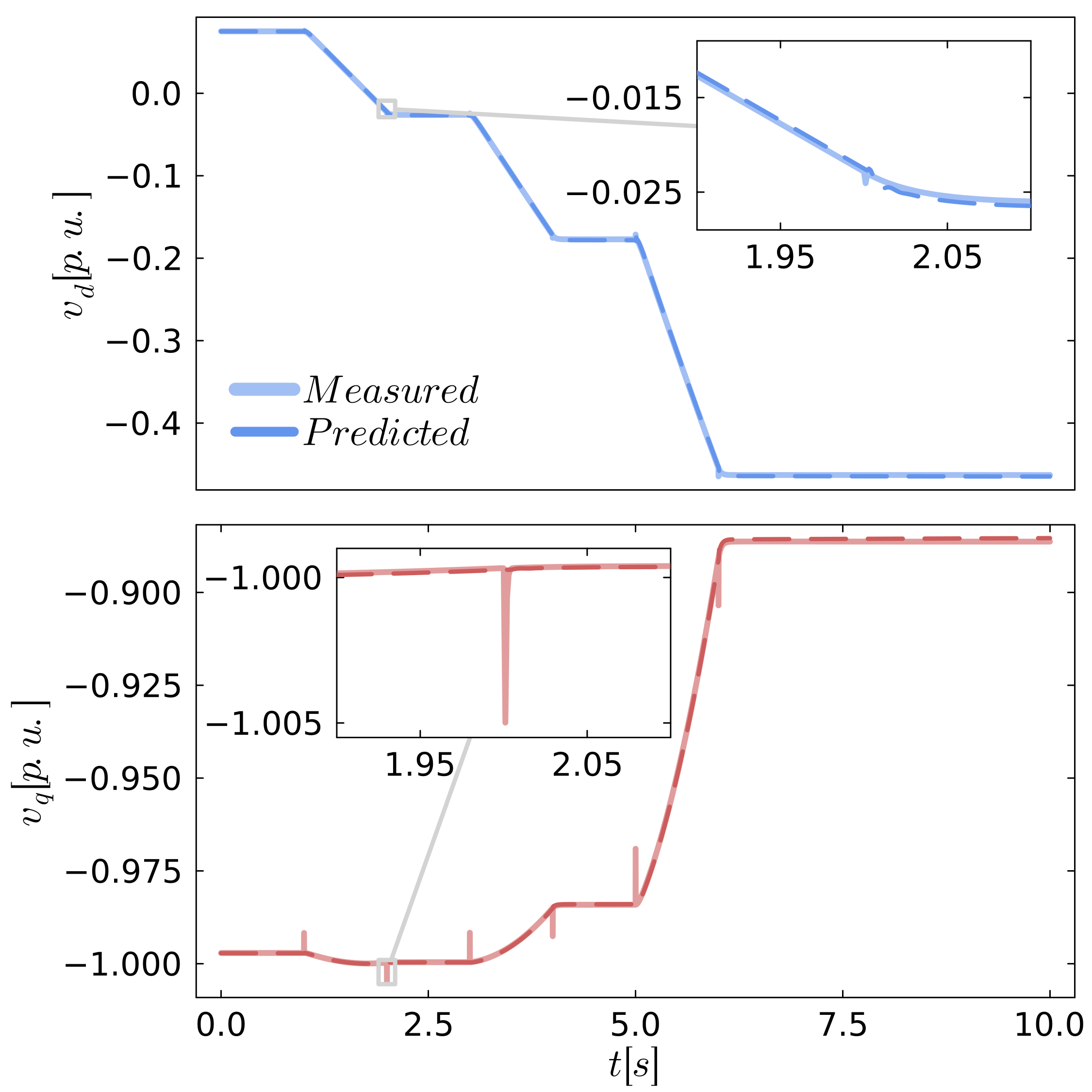}
         \caption{Comparison between the closed-loop normal-form with two internal variables and the dVOC.}
         \label{fig:simulink}
    \end{figure}

    Fig.~\ref{fig:simulink} illustrates a comparison between the closed-loop normal-form with four internal variables and the full dVOC on the out-of-distribution scenario. Notably, both the dVOC and the closed-loop normal-form yield similar voltage transients. The closed-loop normal-form captures the slow transients. However, the fast responses that have been captured by the open-loop model are not fully captured. This phenomenon will be discussed in more detail in the limitations section~\ref{sec:EMT_limitation}.
    In summary, the normal-form has demonstrated good results in modeling the slow dynamics of the dVOC, whether in an open- or closed-loop configuration.
    
    \subsection{Laboratory Experiments}
    For the laboratory experiments, the same procedures as for the simulation data has been used. To emphasize the effectiveness of our approach, we will concentrate on the normal-form with a single internal variable in the following analysis. 

\begin{figure}
        \centering
        \includegraphics[width=\linewidth]{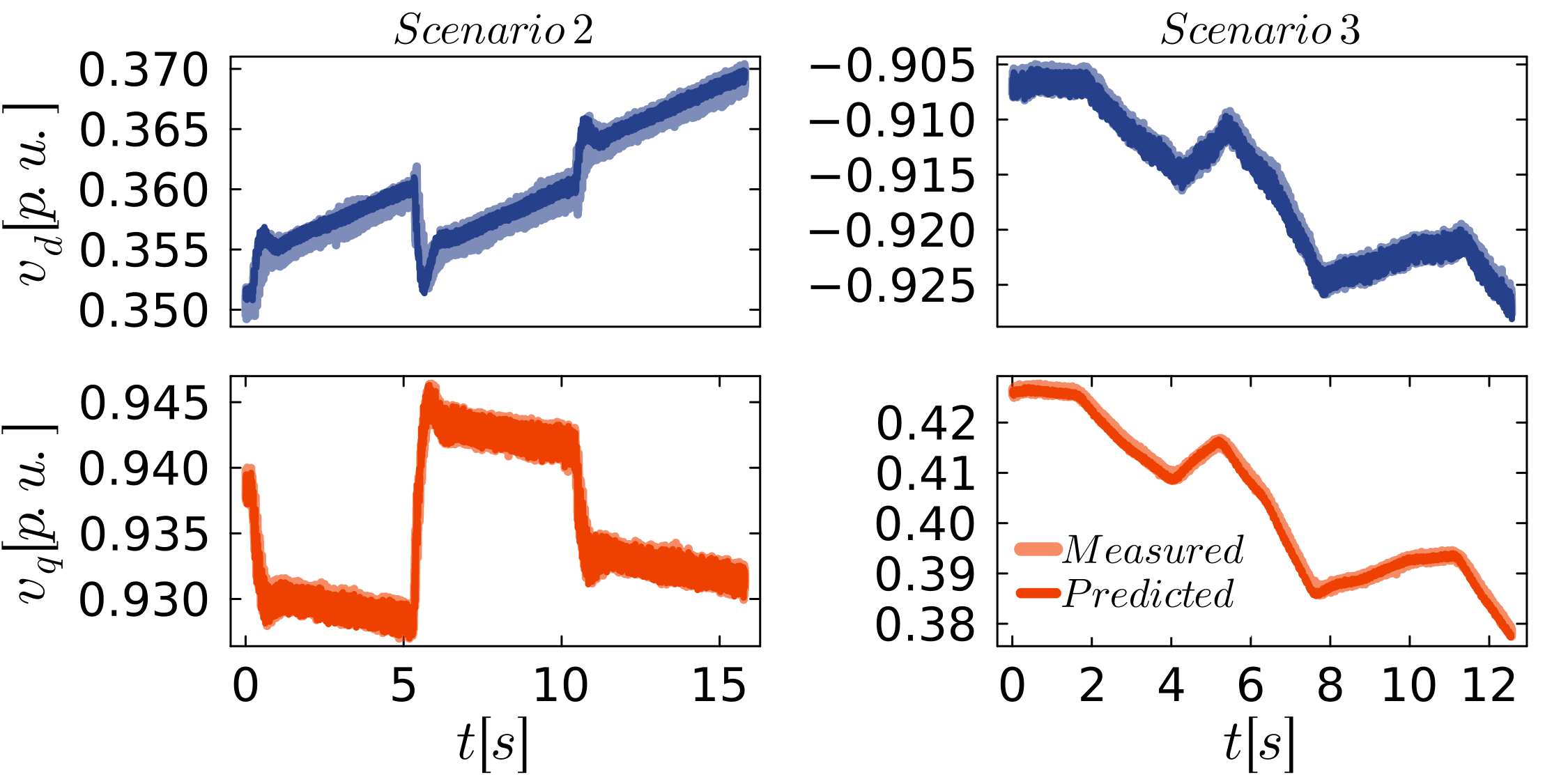}
        \caption{Measured and predicted $\mathrm{dq}$-voltages for the second and third scenario in the test data-set of the droop-controlled inverter.}
        \label{fig:lab_test}
    \end{figure}

    The left pane of Fig.~\ref{fig:lab_test} illustrates the results for the third scenario in the test data-set, where the normal-form successfully captures the voltage dynamics of the inverter. The model performs equally well in the first scenario, which involves step changes in the external grid's voltage magnitude. In the second scenario, involving step changes in the external grid's frequency, the normal-form shows slightly less accuracy, particularly in capturing the harmonics of the $\mathrm{d}$-component, which can be seen in the right pane of Fig.~\ref{fig:lab_test}. Despite this, the overall results remain satisfactory, especially as the collected data is subjected to measurement noise. A more detailed discussion on harmonics will be presented in section~\ref{sec:harmonics}.

\begin{figure}
        \centering
        \includegraphics[width=\linewidth]{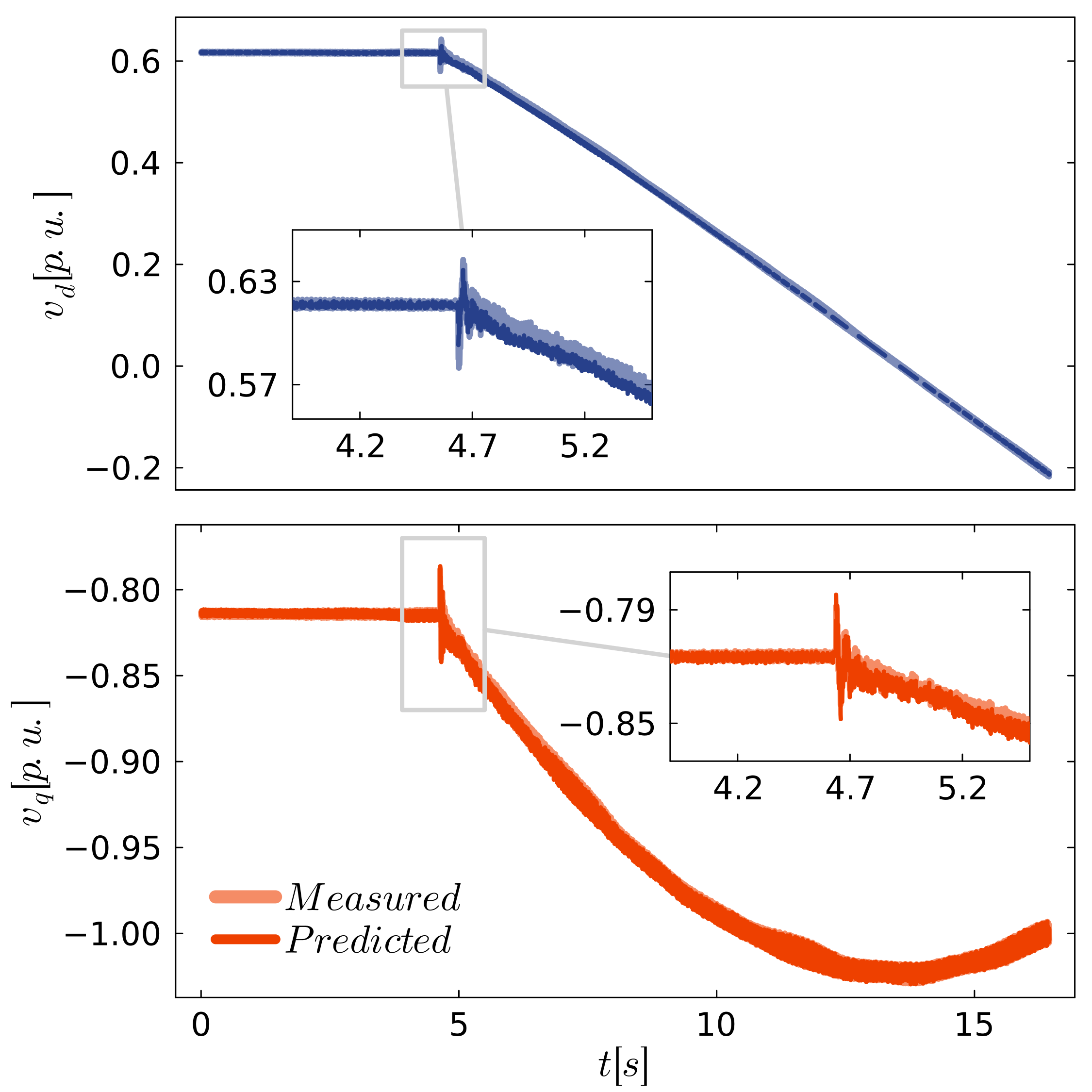}
        \caption{Out-of-distribution performance for the measured and predicted $\dq$-voltage for the droop-controlled inverter. The lens shows the disconnection event, which results in the islanding of the micro-grid.}
        \label{fig:out_of_distribution_lab}
    \end{figure}

    Fig.~\ref{fig:out_of_distribution_lab} depicts the predicted and measured $\mathrm{dq}$-voltage during the islanding process, which is the out-of-distribution task for the laboratory set-up. Remarkably, the normal-form accurately predicts the $\mathrm{dq}$-voltage transients, underscoring its generalization capabilities.
    
\begin{figure}
        \centering
        \includegraphics[width=\linewidth]{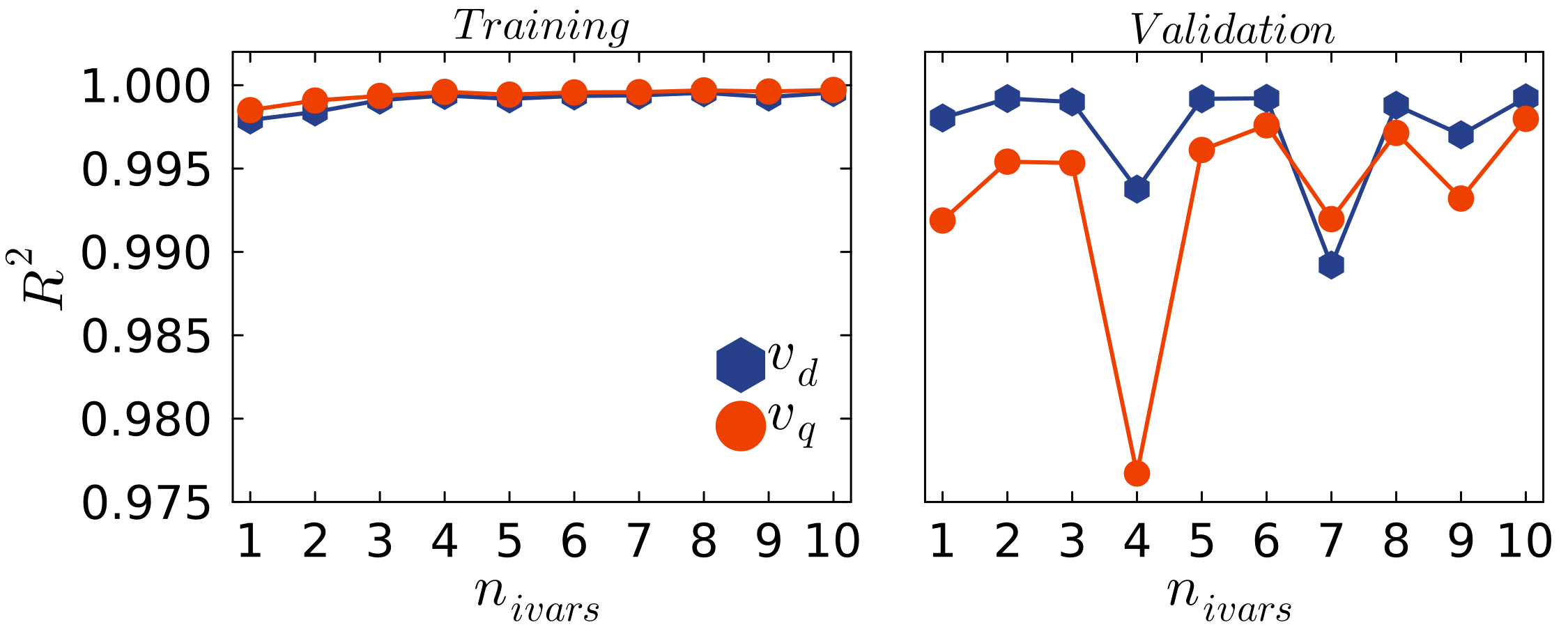}
        \caption{$R^2$ scores with respect to the numbers of internal variables for the droop-controlled inverter. The plots on the left and right show the results for the training and validation data-set respectively.}
        \label{fig:r2_lab}
    \end{figure}

    Fig.~\ref{fig:r2_lab} illustrates the model performance with respect to the number of internal variables. There appears to be minimal dependence on $n_{\mrm{ivars}}$ within the data-set. Notably, only a single internal variable is needed to describe the slow dynamics of the laboratory inverter.
    
    In summary, the normal-form exhibits good results for the laboratory experiments. Since an exact model of the laboratory components is unavailable, testing the closed-loop performance of the inverter was not possible. Once access to a laboratory with a digital twin of all components is available, the closed-loop performance of a laboratory inverter will be studied in subsequent publications.

    \subsection{Limitations}
    \label{sec:limitations}
    In the following, we will focus on two phenomena that are currently not fully captured by either the identification process or the normal-form approach: harmonics and electromagnetic transients.
    
    \subsubsection{Electromagnetic Phenomena}
    \label{sec:EMT_limitation}
    For the dVOC short voltage drops immediately after a load step can be observed in the $\mathrm{dq}$-components, as we can see in Fig.~\ref{fig:EMT_ood}. 

    The load and, thus, the output power of the inverter increases nearly instantaneously. However, the inner current is actively controlled by the cascaded inner loops and takes some time to match the output power. This short imbalance leads to a voltage drop over the capacitor until the inner-loop control takes effect and stabilizes the output voltage.

\begin{figure}
        \centering
        \includegraphics[width=\linewidth]{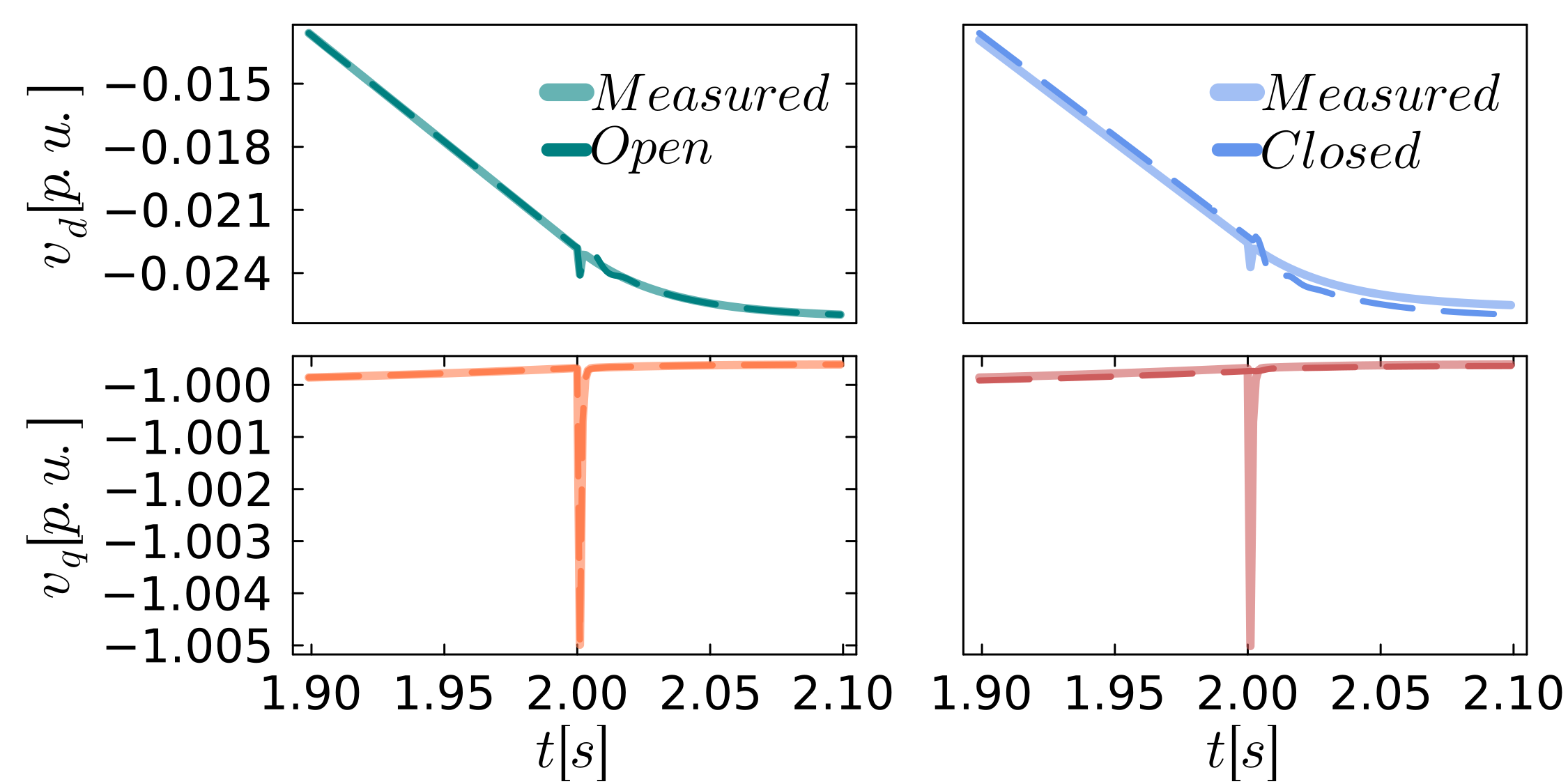}
        \caption{Comparison between open- and closed-loop results.}
        \label{fig:EMT_open_closed_loop}
    \end{figure}

    As shown in Fig.~\ref{fig:EMT_open_closed_loop}, the open-loop normal-form captures these peaks, whereas the closed-loop simulation does not. During those short, highly dynamic events, the feedback between the normal-form output and input in the closed-loop system plays an important role. The normal-form is trained without taking this causality into account and thus cannot predict the correct outcome in this highly dynamic EMT scenario. Hence, this issue does not stem from the normal-form approach itself but rather from the identification process.
     
    To potentially enable the normal-form to capture these electromagnetic transients a \enquote{pseudo-closed-loop} identification could be implemented. A relatively simple update to the open-loop identification, introduced in section~\ref{sec:opt}, would be to close the loop between the predicted complex phase $\cppred$ and the inputs of the normal-form $e$. This means that the inputs are continuously updated during the transients, which can then be added as an additional term to the loss function \eqref{eq:loss}. This approach would allow the normal-form to learn not only the correct output behavior but also how the inputs have manifested. This approach still does not require a model of the ambient power system and thus maintains data-driven. We believe this approach may enable the capturing of EMT effects, however, further research is required. 

    \subsubsection{Harmonic Resonances}
    \label{sec:harmonics}
    This section investigates the capability of the normal-form to model harmonic resonances. In real-world power systems (for example the laboratory experiments), harmonic resonances of the fundamental frequency appear due to the non-linear behavior of certain components, e.g. the discrete switching behavior of the power electronics.
     
    Studies have verified that the extensive use of power-electronic inverters can increase harmonic pollution in power systems \cite{wang_harmonic_2019}. The non-linear dynamics of these inverters significantly contribute to the harmonic response. While linear time-invariant (LTI) models can identify the amplitude and phase shifts introduced by the inverter at each harmonic frequency \cite{wang_harmonic_2019}, they are limited by their inability to generate frequency components not present in the input. Hence, they cannot describe harmonics that are induced by the inverter dynamics themselves. Instead, linear time-periodic (LTP) systems could be employed, as demonstrated in \cite{cecati_ltp_2023}.

    The normal-form, however, is not a pure LTI system. The relation between the non-linear voltage dynamics, given by equation~(\ref{eq:voltage_complex_phase}), and the harmonic response is not well understood. Despite this, we anticipate that the normal-form should exhibit similar capabilities as pure LTI-models. Thus, amplitude- and phase-shifts should be captured by the normal-form while the harmonics induced by inverter dynamics should not be covered. 
  
\begin{figure}
        \centering
        \includegraphics[width=\linewidth]{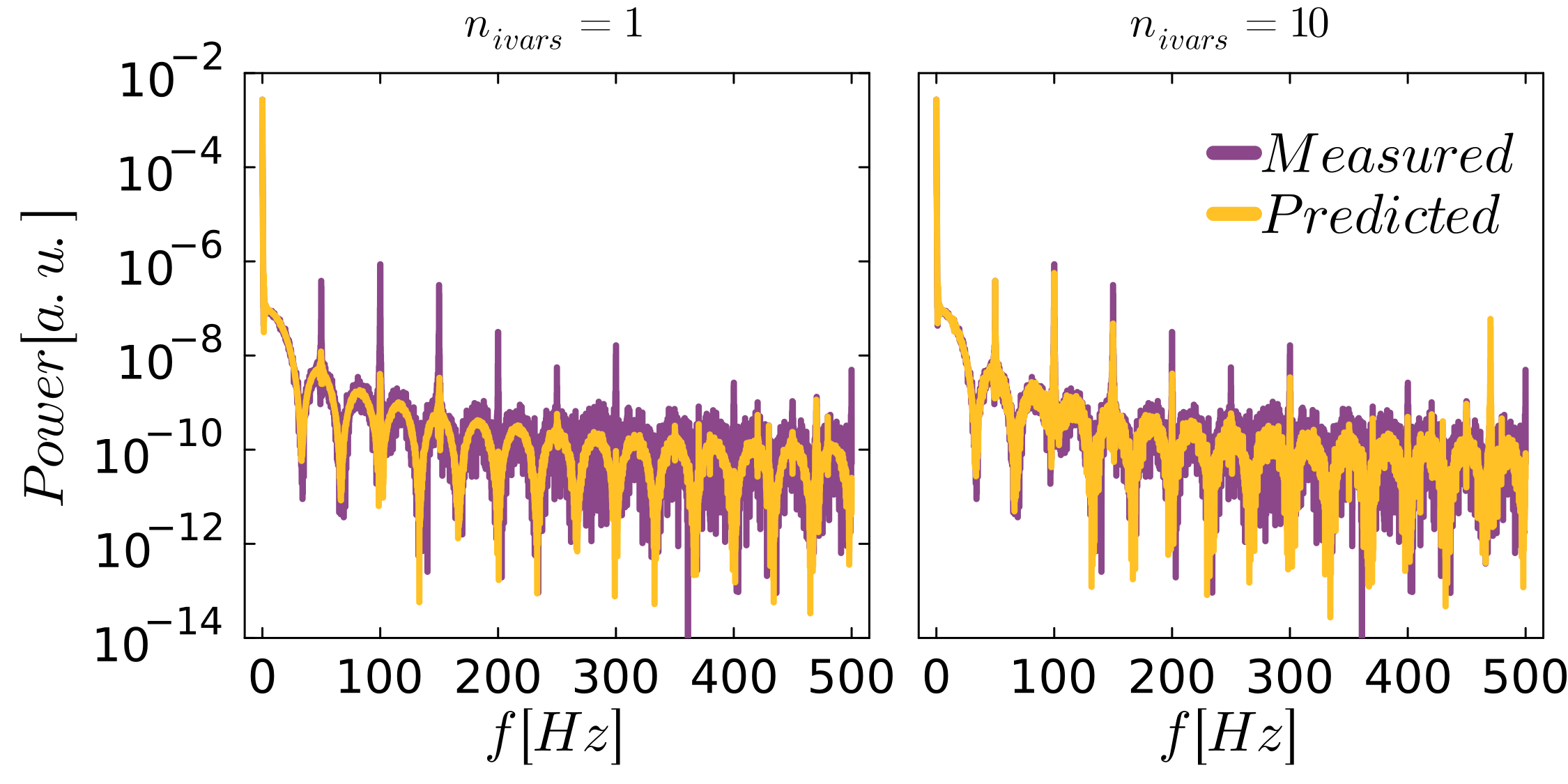}
        \caption{Comparison between the power spectra of the measured and predicted voltage magnitudes for one and ten internal variables.}
        \label{fig:harmonics}
    \end{figure}

    This study examines the harmonics observed in the voltage magnitude $|v|$. Scenario three of the test data-set, which involves periodic changes in the slack voltage magnitude and frequency, has been employed for this. Fig.~\ref{fig:harmonics} compares the power spectra of the laboratory measurements and the normal-form prediction. The measured signal indicates that harmonics of the \SI{50}{Hz} fundamental frequency are clearly present in the data. The high power visible in the low-frequency range is attributed to the slack changes occurring every second.
    
    In Fig.~\ref{fig:harmonics}, the power spectra of the voltage magnitude of a normal-form with one and ten internal variables are compared. It is evident that the model with one internal variable fails to capture harmonics above \SI{150}{Hz} and underestimates the power of all harmonic components. Although the performance improves slightly with ten internal variables, it is clear that merely increasing the number of internal variables is insufficient. 

    In contrast to the EMT-phenomena, the limitation here is given by the modeling approach and not by the identification method. Harmonics introduced by the switching do not satisfy the assumptions underlying the normal-form analysis. Further research on a normal-form approach that includes harmonics is required.

\section{Conclusion and Outlook} 
    Renewable energy sources (RESs) and the power-electronic inverters that integrate them into the grid are becoming increasingly significant in the generation mix. Grid-forming inverters are particularly crucial for the stable operation of RES-dominated systems which necessitates accurate modeling of these devices.

    Despite extensive research, real-life inverters are often treated as black boxes due to undisclosed internal design choices. Existing models, though detailed, are computationally intensive and thus impractical for control-room applications such as dynamic stability assessment tools \cite{matevosyan_grid-forming_2019}. Hence, low-dimensional models derived from data are essential to enable efficient stability assessments.
    
    Inspired by the success of the complex frequency concept \cite{milano_complex_2022}, in this paper we utilize the normal-form model \cite{kogler_normal_2022}, which incorporates complex frequency and phase, to identify low-dimensional models of grid-forming inverters. This theory-driven gray-box model captures the necessary nonlinearities for accurately describing inverter dynamics.

    We have introduced an identification pipeline that is based on the optimization of the $L^2$-norm of the difference between measured and predicted complex phases. This method yielded significantly more robust results compared to using the $\mathrm{dq}$-voltages. Although both complex phase and $\mathrm{dq}$-voltages contain the same information, the complex frequency separates phase angle and magnitude dynamics, which facilitates the identification process. The data collection involves connecting a grid-forming inverter to a stiff voltage source and recording the inverter bus voltages and currents under three different scenarios.
    
    The normal-form model and the accompanying identification process have been validated on two common grid-forming inverter strategies: droop-control \cite{schiffer_conditions_2014} and dispatchable virtual oscillators (dVOC) \cite{colombino_global_2017, seo_dispatchable_2019}. The dVOC was simulated in Simulink, while the droop-controlled inverter was tested in a power hardware-in-the-loop laboratory. 
    
    A normal-form model with four internal variables accurately modeled the dVOC dynamics across all studied data-sets. Both open- and closed-loop normal-form models captured the slow transients of the dVOC, but the closed-loop model struggled with fast EMT phenomena. A \enquote{pseudo-closed-loop} identification, which updates inputs based on predicted outputs, could enhance the model performance in closed-loop simulations without requiring an ambient power system model.
    
    For the droop-controlled inverter, even a normal-form with a single internal variable sufficiently captured the dynamics across all data-sets. However, the normal-form struggled with the representation of harmonics due to its inability to generate frequency components that are not present in the input. Further research is needed to extend the normal-form approach to include harmonic resonances. 
    
    While the results show that the identification remained successful despite the presence of measurement noise in the laboratory, we aim to further investigate noise robustness in a subsequent paper. 
    Furthermore, the description using smooth ordinary differential equations has been found to capture the inverter dynamics successfully. However, in future work, we plan to evaluate the identification performance in scenarios where this modeling assumption may no longer hold, such as in the presence of current limitations. In such cases, the approach may need to be extended to incorporate piecewise modeling frameworks.
    
    The results further demonstrate that the proposed modeling approach can capture large disturbance scenarios, namely, the two out-of-distribution cases studied, load steps, and the islanding experiment. While this was already explored for simple analytical models in \cite{kogler_normal_2022}, our study shows that even significant deviations from the original operating point, such as the disconnect event, are successfully described by the identified models. However, formally proving a range of validity of the identified models remains an open research question.

    In conclusion, the presented results demonstrate the effectiveness of the normal-form approach, and the accompanying identification pipeline, in accurately describing the dynamics of grid-forming inverters. By addressing the challenges of existing computationally intensive models and undisclosed internal design, the normal-form provides a robust, low-dimensional, data-driven alternative. 

\section*{Acknowledgments}
    All authors gratefully acknowledge Land Brandenburg for supporting this project by providing resources on the high-performance computer system at the Potsdam Institute for Climate Impact Research. The work was in parts supported by DFG Grant Number KU 837/39-2 / 360332943, BMWK Project OpPoDyn (FKZ 03EI1071A) and BMWK Projekt MARiE (FKZ 03EI4012A). Anna Büttner acknowledges support from the German Academic Scholarship Foundation.

\section*{Data Availability}
    Both, the data that supports the findings of this study and the source code used to produce the results are openly available on Zenodo \cite{buttner_software_2024}.

    \printbibliography

\end{document}

%% file: tikz/setup.tex
\usepackage{tikz}

\usepackage{tikz}

\usetikzlibrary{shapes.geometric}
\usetikzlibrary{shapes.misc}
\usetikzlibrary{shapes.symbols}
\usetikzlibrary{arrows, arrows.meta, positioning, 3d,calc}
\usetikzlibrary[math, decorations.pathreplacing]
\tikzset{
  >=stealth',
}

\usepackage[straight voltages, american]{circuitikz}
\ctikzset{resistors/scale=0.7,
  capacitors/scale=0.7,
  inductors/scale=0.7,
  instruments/scale=0.5,
}

\makeatletter
\tikzset{self loop/.style =  {to path={
  \pgfextra{}
  [looseness=12,min distance=10mm]
  \tikz@to@curve@path},font=\sffamily\small
  }}
\makeatother

\makeatletter
\tikzset{%
  block/.style      = {draw, thick, rectangle, minimum height = .5cm, minimum width = 1cm, align=center, font=\footnotesize},
  hub/.style        = {draw, circle, fill=black, minimum size=.10cm, inner sep=0.00cm},
  symhub/.style     = {hub, fill=none, draw, inner sep=0.005cm, scale=.5},
  connection/.style = {draw, ->},
  conlabel/.style   = {scale=0.75},
  ede/.style = {circle, fill=lightgray, font=\scriptsize, inner sep=0.05cm, minimum size=.25cm},
  eae/.style = {draw, circle, font=\scriptsize, inner sep=0.05cm, minimum size=.25cm},
}
\makeatother

\makeatletter
\newcommand{\pgfwest}{%
    \pgf@process{\northeast}%
    \pgf@ya=.5\pgf@y%
    \pgf@process{\southwest}%
    \pgf@y=.5\pgf@y%
    \advance\pgf@y by \pgf@ya%
}
\newcommand{\pgfeast}{%
    \pgf@process{\southwest}%
    \pgf@ya=.5\pgf@y%
    \pgf@process{\northeast}%
    \pgf@y=.5\pgf@y%
    \advance\pgf@y by \pgf@ya%
}
\newcommand{\pgfsouth}{%
    \pgf@process{\northeast}%
    \pgf@xa=.5\pgf@x%
    \pgf@process{\southwest}%
    \pgf@x=.5\pgf@x%
    \advance\pgf@x by \pgf@xa%
}
\newcommand{\pgfnorth}{%
    \pgf@process{\southwest}%
    \pgf@xa=.5\pgf@x%
    \pgf@process{\northeast}%
    \pgf@x=.5\pgf@x%
    \advance\pgf@x by \pgf@xa%
}

\long\def\pgfshapeaddanchor#1#2{%
{%
  \def\pgf@sm@shape@name{#1}%
  \let\anchor=\pgf@sh@anchor%
  #2}%
}
\pgfshapeaddanchor{rectangle}{%
  \anchor{west north west}{%
    \pgf@process{\northeast}%
    \pgf@ya=1.5\pgf@y%
    \pgf@process{\southwest}%
    \pgf@y.5\pgf@y%
    \advance\pgf@y by \pgf@ya%
    \pgf@y=.5\pgf@y%
  }%
  \anchor{north north west}{%
    \pgf@process{\southwest}%
    \pgf@xa=1.5\pgf@x%
    \pgf@process{\northeast}%
    \pgf@x.5\pgf@x%
    \advance\pgf@x by \pgf@xa%
    \pgf@x=.5\pgf@x%
  }%
  \anchor{west south  west}{%
    \pgf@process{\northeast}%
    \pgf@ya=.5\pgf@y%
    \pgf@process{\southwest}%
    \pgf@y=1.5\pgf@y%
    \advance\pgf@y by \pgf@ya%
    \pgf@y=.5\pgf@y%
  }%
  \anchor{south south  west}{%
    \pgf@process{\northeast}%
    \pgf@xa=.5\pgf@x%
    \pgf@process{\southwest}%
    \pgf@x=1.5\pgf@x%
    \advance\pgf@x by \pgf@xa%
    \pgf@x=.5\pgf@x%
  }%
  \anchor{west -1}{%
    \pgfwest%
    \advance\pgf@y by -0.5 cm%
  }%
  \anchor{west 1}{%
    \pgfwest%
    \advance\pgf@y by 0.5 cm%
  }%
  \anchor{west A}{%
    \pgfwest%
    \advance\pgf@y by 0.25 cm%
  }
  \anchor{west B}{%
    \pgfwest%
    \advance\pgf@y by -0.25 cm%
  }%
  \anchor{east -1}{%
    \pgfeast%
    \advance\pgf@y by -0.5 cm%
  }%
  \anchor{east 1}{%
    \pgfeast%
    \advance\pgf@y by 0.5 cm%
  }%
  \anchor{east A}{%
    \pgfeast%
    \advance\pgf@y by 0.25 cm%
  }
  \anchor{east B}{%
    \pgfeast%
    \advance\pgf@y by -0.25 cm%
  }%
  \anchor{north -1}{%
    \pgfnorth%
    \advance\pgf@x by -0.5 cm%
  }%
  \anchor{north 1}{%
    \pgfnorth%
    \advance\pgf@x by 0.5 cm%
  }%
  \anchor{north A}{%
    \pgfnorth%
    \advance\pgf@x by -0.25 cm%
  }
  \anchor{north B}{%
    \pgfnorth%
    \advance\pgf@x by 0.25 cm%
  }%
  \anchor{south -1}{%
    \pgfsouth%
    \advance\pgf@x by -0.5 cm%
  }%
  \anchor{south 1}{%
    \pgfsouth%
    \advance\pgf@x by 0.5 cm%
  }%
  \anchor{south A}{%
    \pgfsouth%
    \advance\pgf@x by 0.125 cm%
  }
  \anchor{south B}{%
    \pgfsouth%
    \advance\pgf@x by -0.125 cm%
  }%
}
\makeatother

%% file: tikz/normalform_lti.tex
\begin{tikzpicture}
  \node[block, minimum width=0cm](he){$h^e$};
  \node[block, right=0.75cm of he, minimum height=.75cm, minimum width=0cm](LTI){LTI};
  \node[block, right=0.75cm of LTI, minimum width=0cm](int){$\int\mrm{d}t$};
  \node[block, right=0.75cm of int, minimum width=0cm, minimum height=.75cm,](exp){$\exp(\cdot)$};
  \node[block, right=0.75cm of exp, minimum height=.75cm, minimum width=0cm](PG){\shortstack{\scriptsize power\\\scriptsize grid}};

  \draw[connection](LTI.east) -- (int.west) node[pos=.5, above, conlabel]{$\cf$};
  \draw[connection](int.east) -- (exp.west) node[pos=.5, above, conlabel]{$\cp$};
  \draw[connection](exp.east) -- (PG.west) node[pos=.5, above, conlabel]{$v$};

  \draw[connection](he) -- (LTI) node[conlabel, pos=.5, above]{$e$};
  \draw[connection](-1.75,0) -- (he) node[conlabel, pos=.5, below]{$\Pref, \Qref, \Vref$};
  \draw[connection](PG.east)--++(0.25,0) --++(0,-1.25) -| (he.south B) node[conlabel, pos=0.25,above](){$i$};
  \draw[connection](exp.east)--++(0.375,0) node[hub]{} --++(0,-0.75) -| (he.south A) node[conlabel, pos=0.25,above](){$v$};

\path (he.west) --++ (-0.125,0.5) coordinate (cb1a); 
\path (he.east) --++ ( 0.125,0.5) coordinate (cb1b); 
\draw[decorate, decoration={brace, amplitude=1ex, raise=0ex}] (cb1a) -- (cb1b) node[pos=.5, above=1ex, text width=1.75cm, scale=0.75, align=center] {input\\ nonlinearity};

\path (LTI.west) --++ (-0.125,0.5) coordinate (cb1a); 
\path (int.east) --++ ( 0.125,0.5) coordinate (cb1b); 
\draw[decorate, decoration={brace, amplitude=1ex, raise=0ex}] (cb1a) -- (cb1b) node[pos=.5, above=1ex, text width=1.75cm, scale=0.75, align=center] {linear\\ subsystem};

\path (exp.west) --++ (-0.125,0.5) coordinate (cb1a); 
\path (exp.east) --++ ( 0.125,0.5) coordinate (cb1b); 
\draw[decorate, decoration={brace, amplitude=1ex, raise=0ex}] (cb1a) -- (cb1b) node[pos=.5, above=1ex, text width=1.75cm, scale=0.75, align=center] {output\\ nonlinearity};
\end{tikzpicture}

%% file: tikz/emtsim.tex
\begin{tikzpicture}
  \node[block, minimum width=1cm, minimum height=1.5cm](igbt){};
  \node at (igbt)[rotate=0,text width=1cm,align=center]{\centering\scriptsize{Ideal Voltage Source}};
  \draw (igbt.east 1)
  to [iloop,name=ifsensor,mirror] ++(.5,0)
  to [short,i=\scriptsize $\ifilt$] ++(.5,0)
  to [L] ++(1,0)
  to [short,-*] ++(.25,0) coordinate(csplit){}
  to [short] ++(.75,0)
  to [short] ++(1,0)
  to [iloop,name=igsensor,mirror] ++(.5,0)
  to [short,i=\scriptsize $\iout$,-*] ++(0.5,0) coordinate (terminal);
  \draw (csplit) to [C,-*] (csplit|-igbt.east -1);
  \draw (igbt.east -1) to [short, -*] (igbt.east -1 -| terminal) coordinate (terminalb);
  \draw (csplit)++(0,-0.15)--++(0.65,0) to [rmeter, t=\scriptsize $\uout$, name=vsensor] ++(0,-0.7) to [short] ++(-0.65,0);

  \node(dqabc)[block, below=0.75cm of igbt, minimum width=1cm, minimum height=0.5cm]{};
  \draw[thick, shorten <=0.03cm, shorten >=0.03cm](dqabc.north east) -- (dqabc.south west);
  \path(dqabc.north west) -- (dqabc.south east) node[pos=0.29]{\scriptsize $abc$} node[pos=0.71]{\scriptsize$dq$};

  \draw[connection] (dqabc) -- (igbt);

  \node[block, minimum width=1.5cm, right=.5cm of dqabc, minimum height=0.5cm](CC){\scriptsize Current Ctrl};
  \draw[connection] (CC) -- (dqabc) node[conlabel, above, pos=0.5]{$v^{*}$};

  \node[block, minimum width=1.5cm, right=.5cm of CC, minimum height=0.5cm](VC){\scriptsize Voltage Ctrl};
  \draw[connection] (VC) -- (CC) node[conlabel, above, pos=0.5]{$\ifilt^{*}$};

  \node(abcdq1)[block, above=0.25cm of CC, minimum width=0.75cm, minimum height=0.5cm]{};
  \draw[thick, shorten <=0.03cm, shorten >=0.03cm](abcdq1.north east) -- (abcdq1.south west);
  \path(abcdq1.north west) -- (abcdq1.south east) node[pos=0.29, yshift=0.02cm]{\scriptsize $abc$} node[pos=0.71]{\scriptsize$dq$};

  \node(abcdq2)[block, above=0.25cm of VC, minimum width=0.75cm, minimum height=0.5cm]{};
  \draw[thick, shorten <=0.03cm, shorten >=0.03cm](abcdq2.north east) -- (abcdq2.south west);
  \path(abcdq2.north west) -- (abcdq2.south east) node[pos=0.29, yshift=0.02cm]{\scriptsize $abc$} node[pos=0.71]{\scriptsize$dq$};

  \node at (terminal|-abcdq2.north) (abcdq3)[block, anchor=north east, minimum width=0.75cm, minimum height=0.5cm]{};
  \draw[thick, shorten <=0.03cm, shorten >=0.03cm](abcdq3.north east) -- (abcdq3.south west);
  \path(abcdq3.north west) -- (abcdq3.south east) node[pos=0.29, yshift=0.02cm]{\scriptsize $abc$} node[pos=0.71]{\scriptsize$dq$};

  \draw[connection,<-] (abcdq1.west) --++(-0.15,0) node[conlabel, left]{$\varphi$};
  \draw[connection,<-] (abcdq2.west) --++(-0.15,0) node[conlabel, left]{$\varphi$};
  \draw[connection,<-] (abcdq3.west) --++(-0.15,0) node[conlabel, left]{$\varphi$};

  \draw[connection, preaction={draw, line width=2pt, -, white}] (ifsensor.i) -- (ifsensor.i|-igbt) -| (abcdq1.north);
  \draw[connection] (abcdq1) -- (CC);

  \draw[connection, preaction={draw, line width=2pt, -, white}] (igsensor.i) -- (igsensor.i|-igbt) -| (abcdq3.north);
  \draw[connection] (abcdq2) -- (VC);

  \draw[connection, preaction={draw, line width=2pt, -, white, shorten <=0.5cm}] (vsensor.north) -| (abcdq2.north);

  \node[block, below=0.25cm of CC.south west, minimum width=1cm, minimum height=1cm, anchor=north west](dvoc){\scriptsize dVOC};

  \draw[connection](dvoc.east)++(0,0.25) -| (VC.south) node[conlabel, pos=0.75, left]{$\uout^{*}$};

  \node at (dvoc.south -| terminal)[block, anchor=south east ](power){\scriptsize Power Calc};
  \draw[connection](power.west)--(dvoc.east|-power.west) node[conlabel, anchor=north west, yshift=0cm]{$\Pmeas, \Qmeas$};
  \draw[connection](dvoc.east)--++(0.25,0) node[conlabel, right]{$\varphi$};

  \draw[connection] (abcdq3)--(abcdq3|-power.north);
  \draw[connection] (abcdq2.south)++(0,-0.075) -| (power);

  \draw[connection, <-](dvoc.west)++(0, 0.375) --++(-1,0) node[conlabel, left]{$\Pref$};
  \draw[connection, <-](dvoc.west)++(0, 0.125) --++(-1,0) node[conlabel, left]{$\Qref$};
  \draw[connection, <-](dvoc.west)++(0, -0.125) --++(-1,0) node[conlabel, left]{$\Vref$};
  \draw[connection, <-](dvoc.west)++(0, -0.375) --++(-1,0) node[conlabel, left]{$\wref$};

  \path (terminal)--(terminalb) node [pos=.5, rotate=90]{\scriptsize PCC};;
\end{tikzpicture}

%% file: tikz/control.tex
\begin{tikzpicture}
  \node[block, minimum width=1cm, minimum height=1.5cm](igbt){};
  \node at (igbt) [nigbt, scale=0.6, anchor=centergap]{} ;
  \draw (igbt.east 1)
  to [iloop,name=ifsensor,mirror] ++(.5,0)
  to [short,i=\scriptsize $\ifilt$] ++(.5,0)
  to [L] ++(1,0)
  to [short,-*] ++(.25,0) coordinate(csplit){}
  to [short] ++(.75,0)
  to [L] ++(1,0)
  to [iloop,name=igsensor,mirror] ++(.5,0)
  to [short,i=\scriptsize $\iout$,-*] ++(0.5,0) coordinate (terminal);
  \draw (csplit) to [C,-*] (csplit|-igbt.east -1);
  \draw (igbt.east -1) to [short, -*] (igbt.east -1 -| terminal) coordinate (terminalb);
  \draw (csplit)++(0,-0.15)--++(0.65,0) to [rmeter, t=\scriptsize $\uout$, name=vsensor] ++(0,-0.7) to [short] ++(-0.65,0);

  \node[block, below=0.25cm of igbt, minimum height=0.25cm](pwm){};
  \node at (pwm) {\tiny PWM};
  \draw[connection] (pwm)--(igbt);

  \node(dqabc)[block, below=0.25cm of pwm, minimum width=1cm, minimum height=0.5cm]{};
  \draw[thick, shorten <=0.03cm, shorten >=0.03cm](dqabc.north east) -- (dqabc.south west);
  \path(dqabc.north west) -- (dqabc.south east) node[pos=0.29]{\scriptsize $abc$} node[pos=0.71]{\scriptsize$dq$};

  \draw[connection] (dqabc) -- (pwm);

  \node[block, minimum width=1.5cm, right=.5cm of dqabc, minimum height=0.5cm](CC){\scriptsize Current Ctrl};
  \draw[connection] (CC) -- (dqabc) node[conlabel, above, pos=0.5]{$v^{*}$};

  \node[block, minimum width=1.5cm, right=.5cm of CC, minimum height=0.5cm](VC){\scriptsize Voltage Ctrl};
  \draw[connection] (VC) -- (CC) node[conlabel, above, pos=0.5]{$\ifilt^{*}$};

  \node(abcdq1)[block, above=0.25cm of CC, minimum width=0.75cm, minimum height=0.5cm]{};
  \draw[thick, shorten <=0.03cm, shorten >=0.03cm](abcdq1.north east) -- (abcdq1.south west);
  \path(abcdq1.north west) -- (abcdq1.south east) node[pos=0.29, yshift=0.02cm]{\scriptsize $abc$} node[pos=0.71]{\scriptsize$dq$};

  \node(abcdq2)[block, above=0.25cm of VC, minimum width=0.75cm, minimum height=0.5cm]{};
  \draw[thick, shorten <=0.03cm, shorten >=0.03cm](abcdq2.north east) -- (abcdq2.south west);
  \path(abcdq2.north west) -- (abcdq2.south east) node[pos=0.29, yshift=0.02cm]{\scriptsize $abc$} node[pos=0.71]{\scriptsize$dq$};

  \node at (terminal|-abcdq2.north) (abcdq3)[block, anchor=north east, minimum width=0.75cm, minimum height=0.5cm]{};
  \draw[thick, shorten <=0.03cm, shorten >=0.03cm](abcdq3.north east) -- (abcdq3.south west);
  \path(abcdq3.north west) -- (abcdq3.south east) node[pos=0.29, yshift=0.02cm]{\scriptsize $abc$} node[pos=0.71]{\scriptsize$dq$};

  \draw[connection,<-] (abcdq1.west) --++(-0.15,0) node[conlabel, left]{$\varphi$};
  \draw[connection,<-] (abcdq2.west) --++(-0.15,0) node[conlabel, left]{$\varphi$};
  \draw[connection,<-] (abcdq3.west) --++(-0.15,0) node[conlabel, left]{$\varphi$};

  \draw[connection, preaction={draw, line width=2pt, -, white}] (ifsensor.i) -- (ifsensor.i|-igbt) -| (abcdq1.north);
  \draw[connection] (abcdq1) -- (CC);

  \draw[connection, preaction={draw, line width=2pt, -, white}] (igsensor.i) -- (igsensor.i|-igbt) -| (abcdq3.north);
  \draw[connection] (abcdq2) -- (VC);

  \draw[connection, preaction={draw, line width=2pt, -, white, shorten <=0.5cm}] (vsensor.north) -| (abcdq2.north);

  \node[block, below=0.25cm of CC.south west, minimum width=1cm, minimum height=1cm, anchor=north west](droop){\scriptsize Droop};

  \draw[connection](droop.east)++(0,0.25) -| (VC.south) node [conlabel, pos=0.75, left]{$\uout^{*}$};

  \node at (droop.south -| terminal)[block, anchor=south east ](power){\scriptsize Power Calc};
  \node [block, left=0.25cm of power](lpf){\scriptsize LPF};
  \draw[connection] (power)--(lpf);
  \draw[connection](lpf.west)--(droop.east|-lpf.west) node[conlabel, anchor=north west, yshift=0cm]{$\Pfilt, \Qfilt$};
  \draw[connection](droop.east)--++(0.25,0) node[conlabel, right]{$\varphi$};

  \draw[connection] (abcdq3)--(abcdq3|-power.north);
  \draw[connection] (abcdq2.south)++(0,-0.075) -| (power);

  \draw[connection, <-](droop.west)++(0, 0.375) --++(-1,0) node[conlabel, left]{$\Pref$};
  \draw[connection, <-](droop.west)++(0, 0.125) --++(-1,0) node[conlabel, left]{$\Qref$};
  \draw[connection, <-](droop.west)++(0, -0.125) --++(-1,0) node[conlabel, left]{$\Vref$};
  \draw[connection, <-](droop.west)++(0, -0.375) --++(-1,0) node[conlabel, left]{$\wref$};

  \path (terminal)--(terminalb) node [pos=.5, rotate=90]{\scriptsize PCC};;
\end{tikzpicture}

%% file: tikz/lab_ood.tex
\tikzset{%
  bus/.style = {draw,fill=black, minimum width=1cm, minimum height=0.05cm, inner sep=0.00cm},
}
\begin{tikzpicture}
   \node(b1)[bus] at (0,0){};
   \node(b2)[bus, right=1.5cm of b1]{};
   \node(b3)[bus, right=.75cm of b2]{};
   \node(b4)[bus, right=.75cm of b3]{};

    \path (b1)++(0,-0.5) node[vsourcesinshape, rotate=90, scale=.5](s1){};
    \path (b3)++(0,-0.5) node[vsourcesinshape, rotate=90, scale=.5](s3){};
    \path (b4)++(0,-0.5) node[vsourcesinshape, rotate=90, scale=.5](s4){};
    \draw[] (s1.east) -- (b1);
    \draw[] (s3.east) -- (b3);
    \draw[] (s4.east) -- (b4);
    \draw[-{Straight Barb[length=.2cm]}] (b2)--++(0,-0.6);

    \draw (b1.north) to[short] ++(0,.375) coordinate(b1t);
    \draw (b2.north A) to[short] ++(0,.375) coordinate(b2tA);
    \draw (b2.north B) to[short] ++(0,.375) coordinate(b2tB);
    \draw (b3.north A) to[short] ++(0,.375) coordinate(b3tA);
    \draw (b3.north B) to[short] ++(0,.375) coordinate(b3tB);
    \draw (b4.north) to[short] ++(0,.375) coordinate(b4t);

    \draw (b1t) to[opening switch]++(1.0,0) to[generic, l^=\footnotesize\SI{.5}{\kilo\meter}] (b2tA);
    \draw (b2tB) to[generic, l^=\footnotesize\SI{10}{\kilo\meter}] (b3tA);
    \draw (b3tB) to[generic, l^=\footnotesize\SI{10}{\kilo\meter}] (b4t);

    \node[below=.75cm of b1]{main grid};
    \node[below=.75cm of b2]{load};
    \node[below=.75cm of b3]{droop 1};
    \node[below=.75cm of b4]{droop 2};
\end{tikzpicture}


%% file: tikz/training_flow.tex
\begin{tikzpicture}
  \node[block, minimum width=1.5cm, minimum height=1cm](data){\shortstack{Measured\\Data}};
  \node[block, right=0.75cm of data, minimum width=1cm, minimum height=1cm](ycalc){\shortstack{$\Delta \Pmeas$\\$\Delta \Qmeas$\\$\Delta \numeas$}};
  \node[block, right=.5cm of ycalc, minimum height=1.0cm](LTI){LTI};
  \node[block, right=.75cm of LTI, minimum width=0cm](int){$\int$};
  \node[block, right=0.75cm of int, yshift=-0.375cm, minimum height=1.25cm](loss){Loss\\function};
  \draw[connection](data.east A) -- (data.east A -| ycalc.west) node[pos=.25, above, conlabel]{$\imeas$};
  \draw[connection](data.east B) -- (data.east B -| ycalc.west) node[pos=.25, below, conlabel]{$\Vmeas$};
  \draw[connection](ycalc.east) -- (LTI.west) node[pos=.5, above, conlabel]{$e$};
  \draw[connection](LTI.east) -- (int.west) node[pos=0.5, above, conlabel]{$\cfpred$};
  \draw[connection](int.east) -- (int.east-|loss.west) node[pos=1, above left, conlabel]{$\cppred$};
  \draw[connection] (data.east B)++(0.5,0) node[hub,]{} --++(0,-0.5) coordinate(help){} -- (help-|loss.west) node[above left, conlabel]{$\cpmeas$};
  \draw[connection, <-] (LTI.north)--++(0,0.375) --++(-0.5,0) node[left, font=\scriptsize]{$A^{0}, B^{0}, C^{0}, D^{0}$};
  \draw[connection] (loss.east) --++(0.5,0) node[pos=.5, conlabel, above] {$l$};
\end{tikzpicture}
